\documentclass[aip,
  rsi,amsmath,amssymb,
 reprint,%
 floatfix,
]{revtex4-1}
\usepackage{graphicx}
\usepackage{dcolumn}
\usepackage{bm}
\usepackage{comment}
\usepackage{lipsum}
\usepackage{textcomp}
\usepackage{gensymb}
\usepackage{color,colordvi}

\DeclareMathOperator{\Tr}{Tr}

\begin{document}

\title{Auxetic polymer networks: The role of crosslinking, density and disorder}

\date{\today}

\author{Andrea Ninarello}
\affiliation{CNR Institute of Complex Systems, Uos Sapienza, Piazzale Aldo Moro 2, 00185, Roma, Italy}
\affiliation{Department of Physics, Sapienza University of Rome, Piazzale Aldo Moro 2, 00185 Roma, Italy}

\author{Jos\'e Ruiz-Franco}
\affiliation{Physical Chemistry and Soft Matter, Wageningen University $\&$ Research, Stippeneng 4, 6708WE Wageningen, The Netherlands}
\affiliation{CNR Institute of Complex Systems, Uos Sapienza, Piazzale Aldo Moro 2, 00185, Roma, Italy}

\author{Emanuela Zaccarelli}
\affiliation{CNR Institute of Complex Systems, Uos Sapienza, Piazzale Aldo Moro 2, 00185, Roma, Italy}
\affiliation{Department of Physics, Sapienza University of Rome, Piazzale Aldo Moro 2, 00185 Roma, Italy}

\begin{abstract}
Low-crosslinked polymer networks were recently found to behave auxetically when subjected to small tensions, that is, their Poisson's ratio $\nu$ becomes negative. In addition, for specific state points, numerical simulations revealed that diamond-like networks  reach the limit of mechanical stability, exhibiting values of $\nu = -1$, a condition that we define hyper-auxeticity. This behavior is interesting \textit{per se} for its consequences in material science, but also appealing for fundamental physics because the mechanical instability is accompanied by evidences of criticality.
In this work, we deepen our understanding of this phenomenon by performing a large set of equilibrium and stress-strain simulations in  combination with a phenomenological elasticity theory. The two approaches are found to be in good agreement, confirming the above results. We also extend our investigations to disordered polymer networks and find that the hyper-auxetic behavior also holds in this case, still manifesting a similar critical-like behavior as in the diamond one. Finally, we highlight the role of the number density, that is found to be a relevant control parameter determining the elastic properties of the disordered system.
The validity of the results in disordered conditions paves the way to an experimental investigation of this phenomenon in real systems, such as hydrogels. 


\end{abstract}

\maketitle

\section{Introduction}

Auxeticity is a peculiar characteristic of materials that contract or expand perpendicularly to the direction in which a compressive or extensive strain, respectively, has been applied.~\cite{Lakes1987, evans1991molecular, greaves2011poisson}
Materials showing this behavior are usually characterized by a bulk modulus $K$  significantly smaller than the shear modulus $G$, namely they oppose less resistance to compressive than to shear stress. Indeed, the Poisson's ratio $\nu$ is proportional to $3K-2G$, so that when $K< 2G/3$, negative values of $\nu$ occur. Similarly, if one quantifies responses to uniaxial tensile or compressive deformations through the Young modulus $Y$, auxeticity in the linear regime appears whenever $K < Y/3$. Auxetic materials are an interesting topic not only for fundamental material science, but also because they have potential applications in fields such as medicine~\cite{Bose2012}, sports equipment~\cite{Duncan2018}, and protective clothing~\cite{Tahir2022}.
Furthermore, various materials have been found to exhibit auxetic behavior, including graphene, polymeric foams, textiles, bones, metals.~\cite{Caddock1989, Evans1989, HongHu2011, Gatt2015, Bertoldi2017, Rysaeva2018}

There exist different pathways to obtain auxeticity that have already been documented in literature. They can be divided in two main groups, one of geometrical origin and another of thermodynamical origin. Among those with a geometrical origin, metamaterials raised a particular interest in recent years.~\cite{Larsen1997,Theocaris1997, Bertoldi2017, Hanifpour2018, reid2018auxetic}
On the other hand, auxetic behavior driven by a thermodynamic transition has been known for a long time and it was found in metals such as barium titanate or borophane~\cite{Dong2010, Kou2016} and in polymer gels close to the swelling-deswelling crossover driven by temperature changes.~\cite{Hirotsu1991, Boon2017}

Polymer gels also exhibit auxetic behavior when subjected to a weak tension (negative pressure) at constant temperature, as we found numerically in a recent paper.~\cite{Ninarello2022} In that work, we employed a stress-strain protocol to simulate polymer networks based both on diamond-like and on disordered topology. We discovered that for a given and extremely low-crosslinker concentration, that we were able to reach only for the ordered configurations, the system displays a hyper-auxetic point for which $\nu=-1$. 
Interestingly, this terminal point for mechanical stability is accompanied by critical-like fluctuations of the system volume and a related growing susceptibility, resulting in a vanishing bulk modulus. Such thermodynamic instability thus drives $\nu$ towards its limiting value.
For this reason, we refer to this phenomenon as Hyper-Auxetic Transition (HAT).

In this paper, we will unravel further details concerning the HAT. We begin by focusing on the system topology and show that a hyper-auxetic behavior can also be reached in disordered systems, extending the previous findings for ordered diamond-like networks. This result clarifies that the occurrence of hyper-auxeticity is independent on the system topology, thus suggesting a scenario in which the only reason for an exotic elastic behavior of low density polymer networks is thermodynamic.
Then, we show that despite such unusual behavior, the system elasticity can still be well captured by a phenomenological approach that accounts for small and intermediate deformation response of polymeric materials, the so called Mooney-Rivlin (MR) theory.~\cite{Doghri2013} We thus apply a method, previously employed by some of us to investigate elastic properties of the colloidal version of the currently investigated system, i.e. microgels.~\cite{Rovigatti2019}, to evaluate the elastic moduli of hydrogels via MR and compare it to the numerical results.
Finally, by exploiting the fact that using disordered networks, we can vary both the crosslinker concentration and the number density of the system, which is not possible in diamond networks where the two parameters are coupled, we investigate the role of the system density on the elastic properties. Consequently, we now also grasp the influence of the number density on auxeticity, while in our previous work we only focused on the role of crosslinker concentration.

\section{Models and Methods}

Our computational model of polymer networks relies on bonded repulsive particles, exploiting the interactions first introduced by Kremer and Grest~\cite{Grest1986, Kremer1990} and extensively employed to simulate numerous polymer systems, from chains to networks.~\cite{Duering1992, Duering1994, Kenkare1998, Auhl2003, Everaers2004, Lang2013}
All monomers interact through a repulsive component  consisting of a Weeks-Chandler-Andersen (WCA) potential:
\begin{equation}
\label{eq:WCA}
V_{WCA}\left( r \right )=\left\{\begin{matrix}
 4\epsilon\left[\left(\frac{\sigma}{r} \right )^{12}-\left(\frac{\sigma}{r} \right )^{6} \right ]+\epsilon &\ \ \  if \ \ \  r \leq 2^{1/6}\sigma\\ 
0 &\ \ \  if \ \ \  r > 2^{1/6}\sigma
\end{matrix}\right.
\end{equation}
where $\epsilon$ controls the energy scale and $\sigma$ is the monomer diameter that set the unit length. In our simulations, the unit time is defined as $\tau = \sqrt{\frac{m\sigma^2}{\epsilon}}$. Chemical bonds between particles are modeled with a FENE potential:
\begin{equation}
V_{FENE}(r)=-\epsilon k_F R_0^2 ln \left[ 1 - \left(\frac{r}{R_0\sigma}\right) \right] \  \ \text{if} \ r< R_0\sigma
\end{equation}
This model is reliably reproducing the statistical and dynamical properties of polymer chains,~\cite{Duering1991,Duering1992,Duering1994} allowing as well the simulation of systems featuring up to hundred thousands of particles in a reasonable time. 

We both employ ordered and disordered networks. The ordered model consist of a diamond-like structure made up of unitary cells having chains of equal length connected through crosslinkers placed at the lattice atom position. We simulate systems with $8$ unitary cells for different crosslinker concentrations $c=\frac{N_c}{N}$ given by the ratio between the number of crosslinkers $N_c$ with respect to the total number of particles $N$. As a consequence of the topology, the chain length is $l=\frac{1-c}{2c}$.
On the other hand, disordered networks are obtained through a recently introduced protocol based on self-assembly of patchy particles~\cite{Gnan2017}, allowing for the computational fabrication of networks with an exponential strand length distribution, in agreement with the Flory-Rehner theory and analogous to that of experimental systems.~\cite{Sorichetti2021,Sorichetti2023} The method relies on simulation of patchy particles with two or four attractive patches mimicking respectively monomers and crosslinkers. This is performed in a given volume, which determines the resulting number density of the network.  We perform molecular dynamics simulations of this system at low temperature and density, so that the particles assemble as previously theoretically predicted and computationally observed.~\cite{Sciortino2007,Sciortino2011} We wait for the bonds to be almost completely satisfied, with bonding percentage around $p_b \approx 99.9\%$ and then we remove the few particles that do not belong to the main cluster (less than $4\%$ of the total) and the few ($\lesssim1\%$ of all monomers) pertaining to dangling ends, in order to have a fully-bonded network.

Assembly simulations are carried out using the oxDNA simulation package,~\cite{Rovigatti2014, Poppleton2023} and then  we replace the patchy interactions by the Kremer-Grest potential to obtain a chemically bound system. We self-assemble the network by fixing the temperature at $T=0.03$ and by varying the initial number density, in order to control the final density. Further details on the correspondence between initial and final density of the network can be found in Refs.~\citep{Sorichetti2021, Sorichetti2023}
Hereafter, we only refer to the final density leaving aside the initial assembly density.
It is important to note that spherical networks assembled in this way are able to structurally and topologically reproduce the experimental behavior of microgel particles, the colloidal counterpart of hydrogel systems, in a quantitative way as extensively discussed in previous literature.~\cite{Gnan2017, Ninarello2019} Therefore, we expect the disordered configurations of the hydrogels to be representative of experimental hydrogels.

Values of total number of particles $N$, number of crosslinkers $N_{c}$, crosslinker concentration $c$ and number density of the network equilibrated at $P=0$ are reported in Table~\ref{tab:HLsys} for the systems investigated in the following, where we also include values referred to the ordered systems for completeness. 

\begin{table}
\begin{center}
\begin{tabular}[c]{l | c | c | c | r}
\hline
id 		& N		&	$N_c$	&	c ($\%$)	& $\rho(P=0)$ 	\\ \hline
Diam   	&18240	&	64		& 0.35  & 0.0218	\\
Diso 	&14216	&	179		& 1.26  & 0.0203	\\
$L_3$ 	&4881	&	148		& 3.03	& 0.103		\\
$L_7$	&4882	&	371		& 7.60	& 0.103  	\\
$H_3$	&4950	&	149		& 3.01 	& 0.192		\\
$H_7$ 	&4951	&	373		& 7.53	& 0.187		\\
$D_3$	&7128	&	216		& 3.03	& 0.0966	\\
$D_7$ 	&6656	&	512		& 7.69	& 0.187		\\
\hline
\end{tabular}
\caption{\label{tab:HLsys} Name, total number of particles $N$, number of crosslinkers $N_c$, crosslinker concentration $c$, and density at $P=0$ for all the considered systems.}
\end{center}
\end{table}

For both ordered and disordered systems we perform NPT simulations at negative pressures using LAMMPS simulation package~\cite{Plimpton1995} with a Nosé-Hoover thermostat and barostat.
Temperature is set to $1.0$ and is measured in units of energy, i.e. fixing $k_B = 1$, where $k_B$ is the Boltzmann constant. Simulations at different pressures have been performed using a timestep $\delta t = 0.003\tau$. We perform, independently, equilibrium and strain-stress simulations with the aim of investigating elastic properties.

Equilibrium simulations are performed both to compute the bulk modulus straight from volume fluctuations, given that $K=k_BT\frac{\langle V \rangle}{\langle V^2 \rangle -\langle V \rangle^2}$, as well as to apply the Mooney Rivlin theory.  This framework was previously employed to compute elastic properties of microgels in bulk, both standard~\cite{Rovigatti2019} and composite~\cite{RivasBarbosa2022}, and at an interface~\cite{Camerin2020}. The theory relies on computing the Green-Lagrange strain tensor ${\tilde{C}}=\tilde{F}^T\cdot \tilde{F}$, providing a measure of the local deformation from the deformation gradient tensor $\tilde{F}$ with respect to a reference configuration.
Consequently one is able to compute three strain invariants:
\begin{align}
J &= \sqrt{\det \tilde{C}} \\
I_1 &= \Tr(\tilde{C})J^{-2/3} \\
I_2 &= \frac{1}{2}[\Tr^2(\tilde{C})-\Tr(\tilde{C}^2)]J^{-4/3}
\end{align}
where $J$ accounts for volume changes and $I_{1,2}$ for variations of shape at constant volume. In the undeformed reference configuration the strain invariant values are by definition $J_{ref} = 1, I_{1,ref} = I_{2,ref} = 3$. From the simulations, we find that $I_1 = I_2=I$ as in Ref.~\citep{Rovigatti2019}.
The Mooney-Rivlin theory relies on a phenomenological expression of the stored elastic energy that can be written in terms of strain invariants:~\cite{Little2023}
\begin{align}
U(J, I)&= U_0 + W(J) + W(I)  \nonumber  \\
=U_0 + &V \left[ \frac{K}{2}\frac{(J-1)^2}{J} + (C_{10} + C_{01}) (I + 3) \right].
\label{eq:energyMR}
\end{align}

The $J$ factor at the denominator in the first part of this equation is usually neglected since $J\approx 1$ for configurations close to the reference state. Here, we include it for completeness. Following Ref.~\cite{Aggarwal2016, Aggarwal2018} one can approximate the $W$  function with the potential of mean force obtained from the probability distribution functions:
$W(X) = -k_B T \ln P(X)$ where $X=J,I$. Then, by fitting this probability distribution to  a function of the form $M_J(J-1)^2/J + D_0$ and $M_I I^3 + D_1$ respectively for $J,I$, where $D_0,D_1$ are fitting parameters, one can obtain two elastic moduli through the relations: $K=\frac{2M_J}{V}$ and $G=\frac{2 M_I}{V}$. The other moduli can be subsequently computed using the relations from linear elasticity theory.\cite{landau_theory_2009}
In case of finite-size objects presenting anisotropy, such as macromolecules or microgels, estimating the reference configuration and the corresponding deformation generally implies  some kind of coarse-graining on the system geometry.~\cite{Aggarwal2016, Aggarwal2018, Rovigatti2019} For infinite systems like the ones we deal with in this paper, the situation can be simplified, as one can take as the reference state a box with sides $\langle L_x \rangle$, $\langle L_y \rangle$, $\langle L_z \rangle$ corresponding to average side lengths and thus quantify deformation in terms of the deviation with respect to this reference configuration due to thermal excitations. This allows us to write the strain invariants as:
\begin{align}
J&=\frac{L_x}{\langle L_x \rangle}\frac{L_y}{\langle L_y \rangle}\frac{L_z}{\langle L_z \rangle}\\
I&=\left(\frac{L_x}{\langle L_x \rangle} + \frac{L_y}{\langle L_y \rangle} + \frac{L_z}{\langle L_z \rangle}\right)J^{-2/3}.
\label{eq:strain}
\end{align}

We also perform strain-stress simulations in order to simultaneously calculate $Y$ and $\nu$ as in Ref.~\cite{Ninarello2022}.
The equilibrated configuration is thus subjected to a longitudinal extensional strain $\lambda_\parallel= \frac{L_\parallel-L^0_\parallel}{L^0_\parallel}$, where $L^0_\parallel$ and $L_\parallel$ are the initial and the final box length respectively along the axis of deformation. We employ values of the deformation within the interval $\lambda_\parallel \in [0,0.3]$ at a fixed strain rate $\dot\lambda = 0.01 \tau^{-1}$. For these values the response of the system is in the linear elastic regime.
We allow the box to readjust independently in the transversal directions in order to obtain an average constant $P$.
We then calculate the stress along the deformation axis,  $\sigma_\parallel$,  from the virial stress tensor averaged over $10^6\tau$ and consequently the Young modulus,  $Y=\frac{\sigma_\parallel}{\lambda_\parallel}$.
Simultaneously, we obtain $\nu$ from transversal fluctuations, using the expression $\nu=\frac{-\partial \lambda_\perp}{\partial \lambda_\parallel}$, where $\lambda_\perp = \frac{\lambda_2 + \lambda_3}{2}$ and $\lambda_{2,3}$ are the components of the strain perpendicular to the deformation axis, respectively. Each configuration is deformed in the three spatial directions independently and results are averaged over them. For each spatial direction $20$ independent deformations are performed starting from the same structural configuration, but with different velocities extracted from a Maxwell-Boltzmann distribution.

\section{Results}
\subsection{Hyper-auxetic transition in disordered networks under tension}

\begin{figure}
\includegraphics[width=0.48\textwidth]{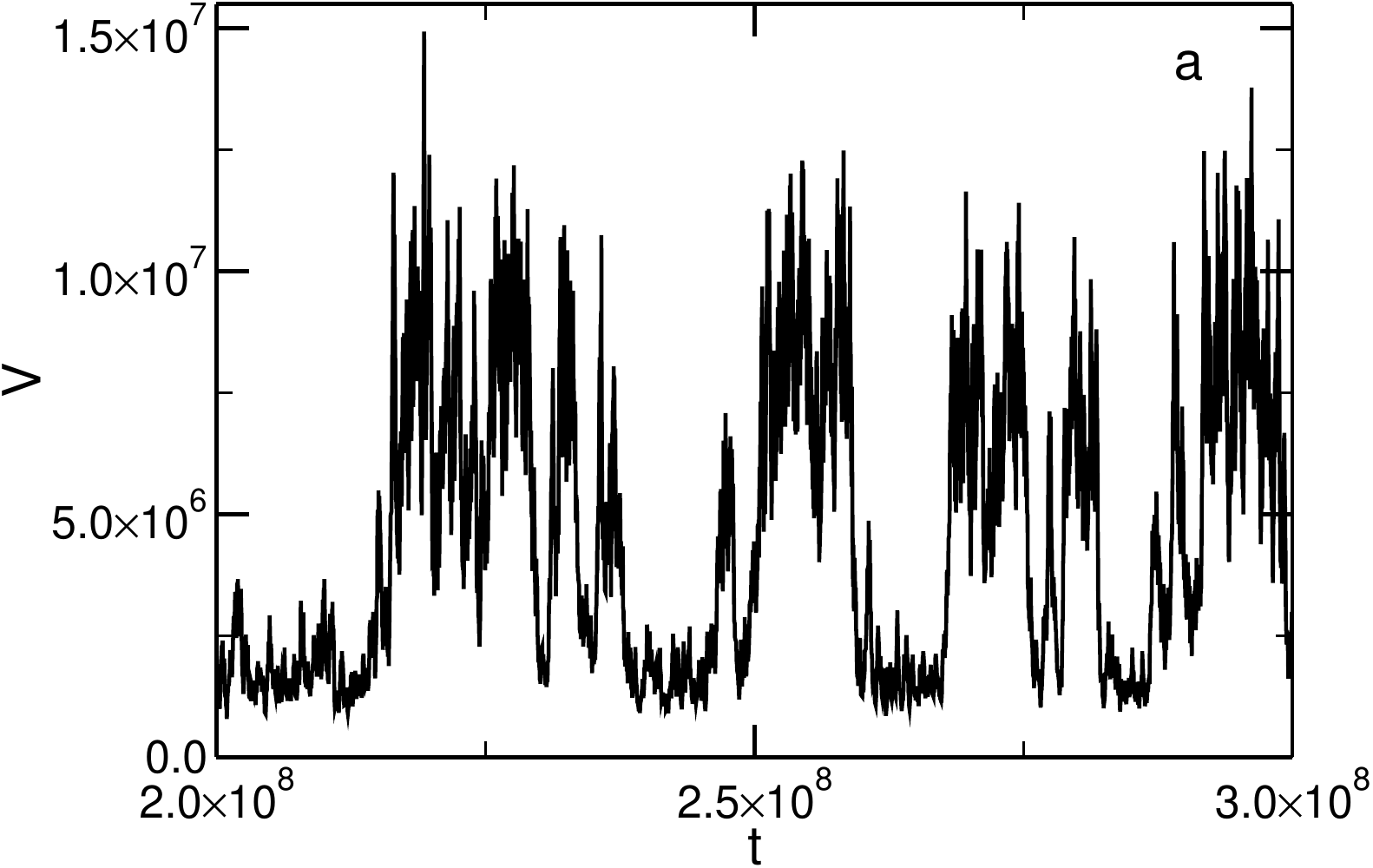}
\includegraphics[width=0.43\textwidth]{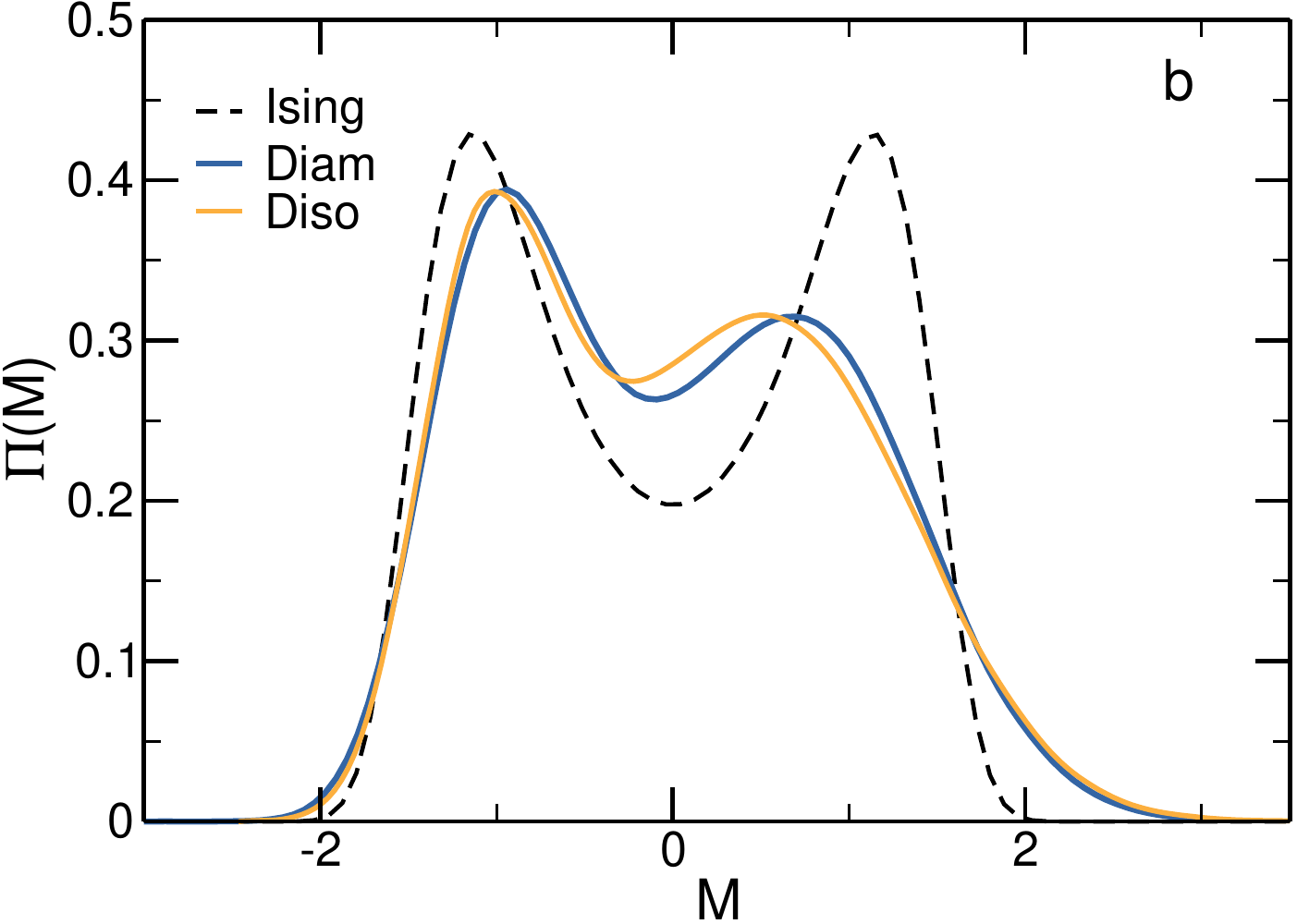}
\centering
\caption{(a) Volume time series for the disordered system with $c=1\%$ at $P=4.675\times 10^{-5}$. (b) Probability distribution of the order parameter $M=\rho+s e_{nb}$, as discussed in the text,  for the disordered system at $c=1\%$ (orange line) $P=4.675\times 10^{-5}$ and the diamond $c=0.35\%$ system at $-8.19\times 10^{-5}$ (blue line). The black dashed line show the theoretical expectation for an Ising transition.}
\label{fig:vol_crit}
\end{figure}

We start by investigating whether the HAT phenomenology is somehow linked to geometry. As discussed earlier, in some cases geometry has been identified as an important factor inducing auxetic behavior. Being this a consequence of material design, such as in metamaterials, or spontaneously happening, such as in polymer foams, a particular topology is usually considered to be the main ingredient controlling how a given system responds to external perturbation.
With the aim of understanding whether this is the case also for polymer networks, we now focus on disordered systems. Already in Ref.\cite{Ninarello2022}, we detected the occurrence of negative values of the Poisson's ratio at small negative pressures, but we observed clear signs of hyper-auxeticity and critical-like fluctuations only for ordered (diamond-like) networks.

\begin{figure}
\centering
\includegraphics[width=0.30\textwidth ]{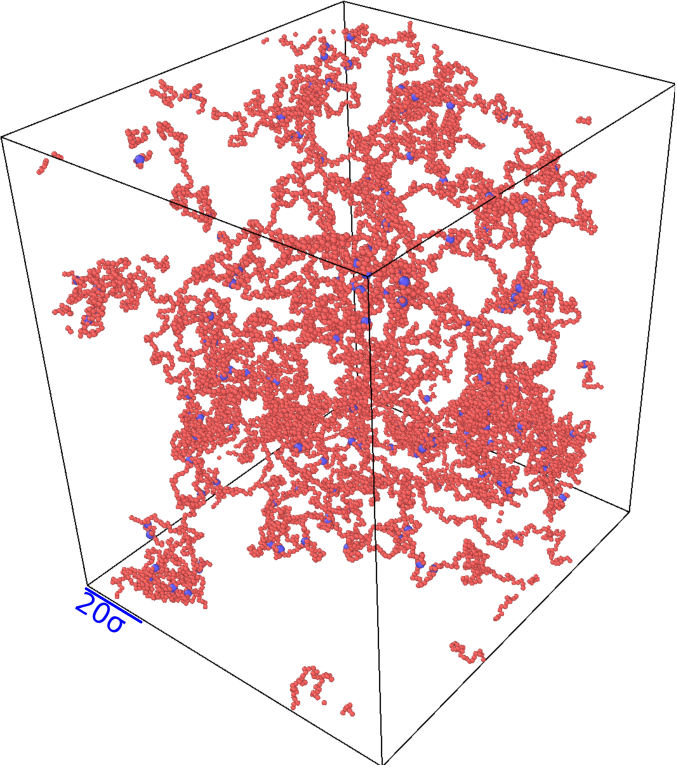}\\
\vspace{0.4cm}
\centering
\includegraphics[width=0.38\textwidth ]{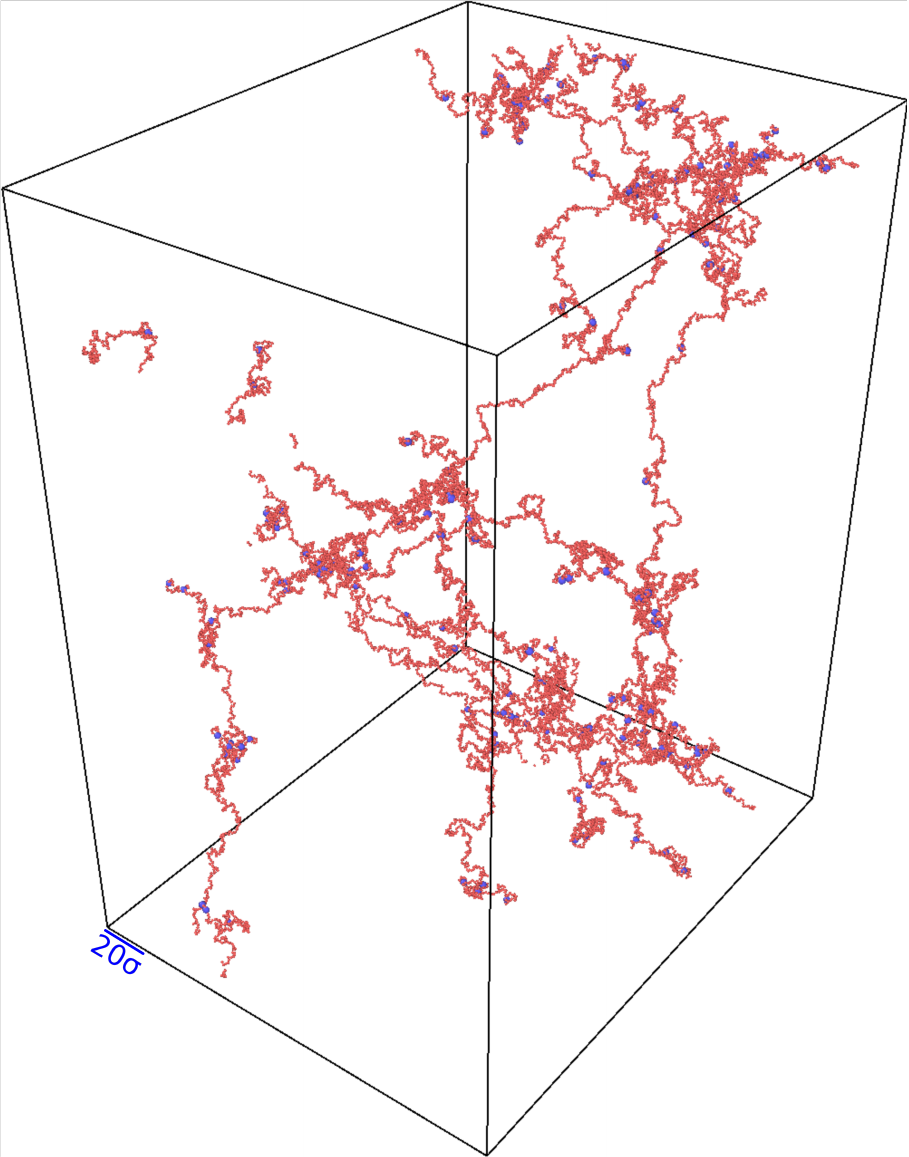}
\caption{Snapshots of a compressed (top) and of an expanded system (bottom) for the \textit{Diso} system at $P=-4.675\times10^{-5}$ taken during equilibrium simulations. Box dimensions are rescaled for representation purpose and relative scales are indicated in the figure. Blue monomers represent the crosslinkers, while red ones are all the other monomers.}
\label{fig:snaps}
\end{figure}

Here we expand the previous results and consider disordered networks of even smaller densities with respect to previously investigated systems. We note that the assembly of such networks is computationally cumbersome and can take weeks of GPU-accelerated computing.
In addition, the nominal assembly density does not strictly coincide with the final density of the network, as already noted in Ref.\cite{Sorichetti2023}. Hence, we prepared a few realizations of low-density disordered network with crosslinker concentration $c=1\%$ and studied their equilibrium behavior.  Among them, we found a realization, here referred to as \textit{Diso} featuring structural values as reported in Table~\ref{tab:HLsys} that clearly displays the occurrence of a HAT at a given value of the pressure, in full analogy to the case of ordered networks. From a purely thermodynamical viewpoint this consists in critical-like fluctuations of the volume, that are reported in Fig.~\ref{fig:vol_crit}(a), covering roughly two orders of magnitude of variation in $V$. The corresponding
 snapshots,  respectively in a compressed and in an expanded state taken at the same critical pressure $P_c=-4.675\times10^{-5}$, are shown in Fig.~\ref{fig:snaps}.

In order to rationalize these qualitative observations, we then calculate the probability distribution of the order parameter $M=\rho + s\epsilon_{nb}$, which combines the density and the energy of non bonded particles through a mixing parameter $s$. As shown in Ref.~\cite{Ninarello2022}, this generalizes the gas-liquid order parameter to the present polymer networks. Indeed, in standard gas-liquid transition, the order parameter includes the total energy, but as previously shown~\cite{Ninarello2022}, this is not explicitly involved in the transition, because it is largely dominated by the bonding energy between connected monomers. When we subtract such bonding contribution, we find that the energy of the non-bonded particles, i.e., the excluded volume contribution coming from the WCA potential, correlates with the density. After calculating $\Pi(M)$, we then rescale it through its mean and standard deviation and compare it to the Ising distribution in Fig.~\ref{fig:vol_crit}(b). We find that $\Pi(M)$ resembles the Ising one although not completely matching the expected universal behavior.  Clear differences, particularly in the high-density (right) peak are present. For completeness, we also compare the \textit{Diso} results with those previously obtained for the ordered system with $c=0.35\%$, here referred  as $Diam$. Remarkably, we find a very good agreement between the two networks, despite them being intrinsically very different both in terms of topology and in number of crosslinks. Indeed, they are much more similar to each other than to the Ising reference curve. 

\begin{figure*}[t]
\includegraphics[width=0.325\textwidth]{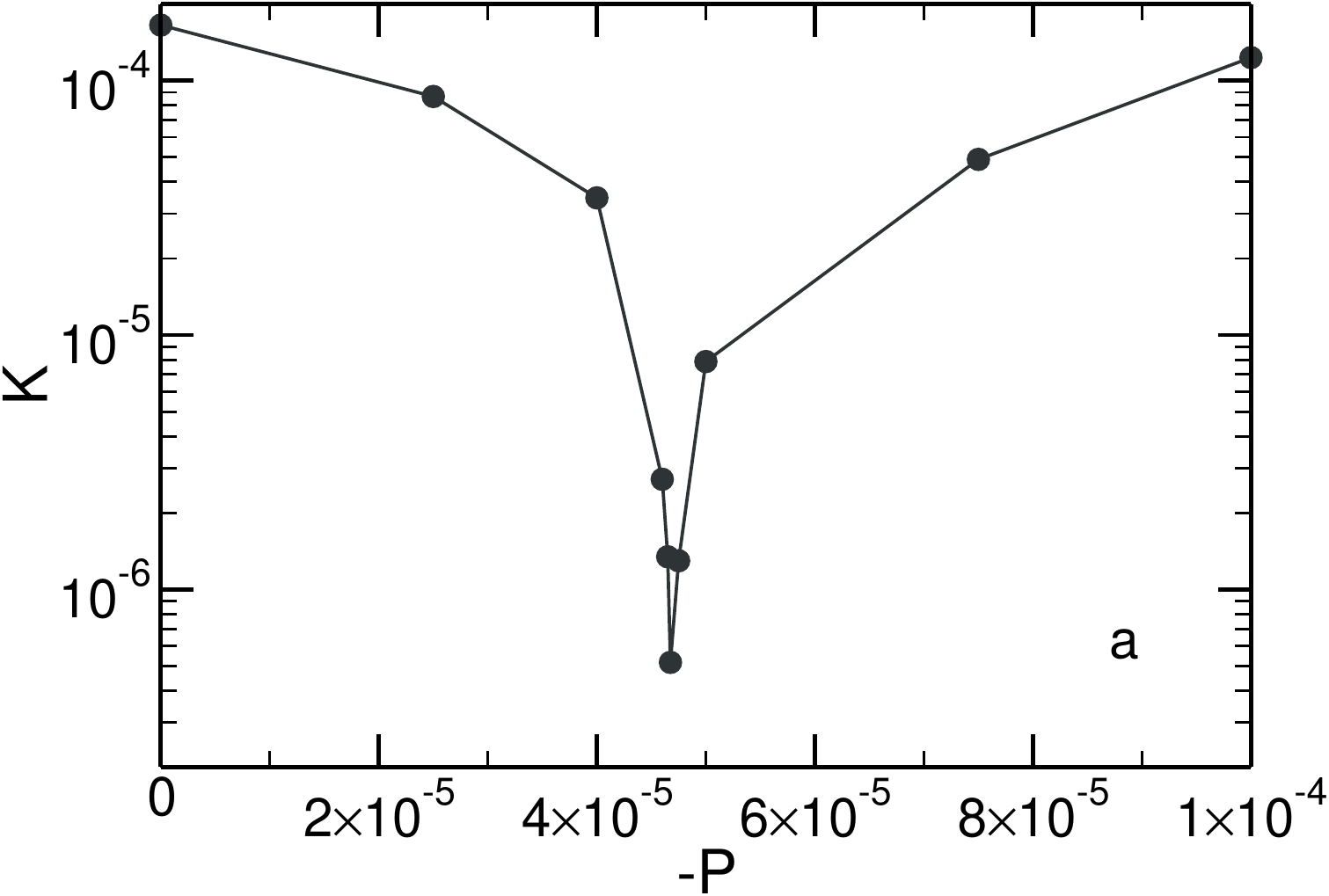}
\includegraphics[width=0.325\textwidth]{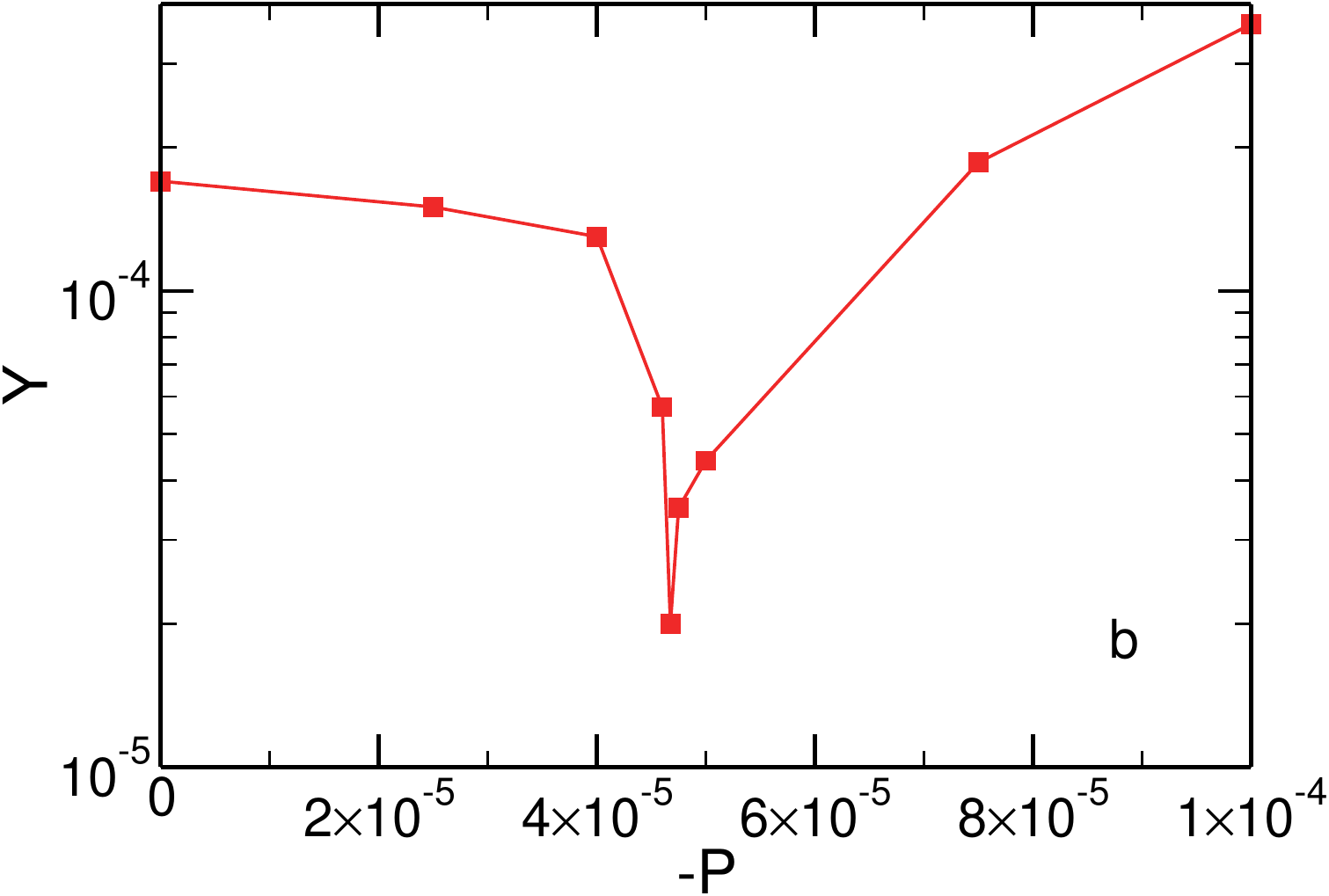}
\includegraphics[width=0.325\textwidth]{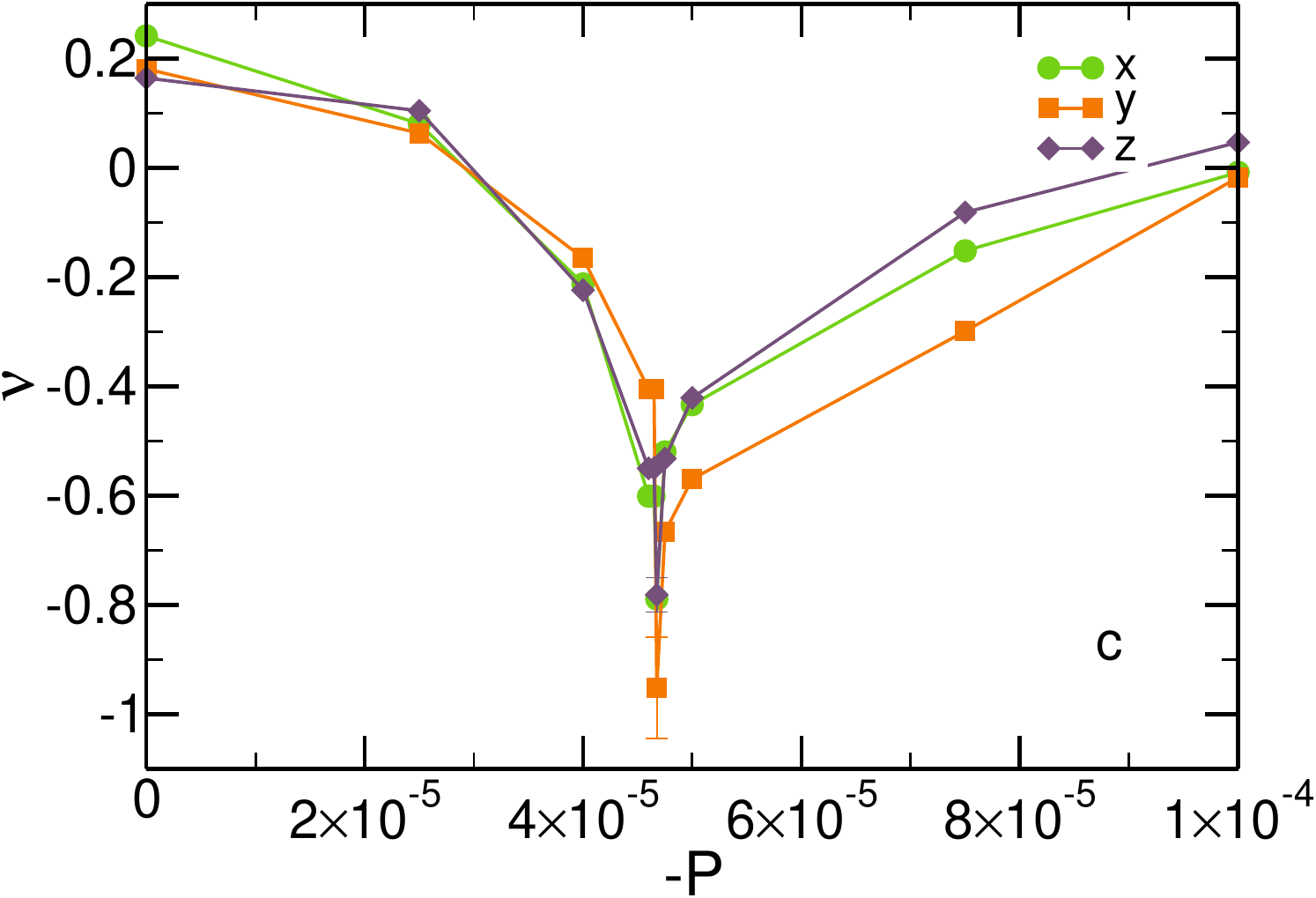}
\centering
\caption{Elastic moduli as a function of $-P$ for the disordered $c=1\%$ (\textit{Diso}) system: (a) bulk modulus, (b) Young modulus and  (c) Poisson's ratio. For the latter, the three different curves are related to three different strain orientations following the $x,y,z$ axis.}
\label{fig:diso_trans}
\end{figure*}

Carrying on the analogy with the HAT observation in ordered systems, we also calculate the elastic properties of the \textit{Diso} network, performing both equilibrium and strain-stress simulations around $P_c$. Namely, we first compute the bulk modulus $K$ from equilibrium fluctuations, which is reported in Fig.~\ref{fig:diso_trans}(a) as a function of pressure. We observe that this observable has a sharp minimum at $P_c$,  decreasing by more than two orders of magnitude with respect to its value at $P=0$.

Next, we calculate the elastic moduli with strain-stress simulations as described in Methods.  The resulting Young modulus $Y$ and  Poisson's ratio $\nu$ are reported in Fig.~\ref{fig:diso_trans}(b) and (c), respectively.
We observe a deep minimum in $Y$, that also decreases by roughly one order of magnitude, again similarly to what found for the diamond system.
As previously discussed, a necessary condition for auxeticity is that the response to isotropic compression should be much weaker as compared to strain solicitations, that is precisely what we observe around $P_c$, where negative values of $\nu$ occur. To rationalize this behavior we also recall that $\nu =\frac{3K-Y}{6K}$. However,  differently from what seen in ordered systems, we find the elastic response significantly varies depending on the strain direction. This effect, that can be ascribed to the system heterogeneity, results in values of the Poisson's ratio that are close to the hyper-auxetic scenario ($\nu\sim-1$), within the numerical resolution of the present simulations,  only in one direction (specifically the $y$-axis).  For $\nu$ measured along the $x$-axis and $z$-axis, we found the minimum value of $\nu$ around $\nu=-0.8$. 
The anisotropic response of the network indicates the presence of strong heterogeneities in the system, which results from our assembly process and the corresponding intrinsic disorder of each generated topology, that, as it will discussed in the following Sec.~\ref{sec:density}, relates to chain length distribution. Hence, an average over the disorder would be as profitable as challenging to perform, given the difficulty to assemble the network at very low investigated connectivities.

We note on passing that the density at $P=0$ of the $Diam$ and $Diso$ systems undergoing HAT are quite similar, as reported in Table~\ref{tab:HLsys}. This observation makes us ponder whether the crosslinker concentration is the true main control parameter for the transition. To this aim we focus on the importance of the number density in the determination of the elastic properties, as discussed in section~\ref{sec:density}.

Finally, up to now, we have been discussing results for $Y$ and $\nu$ obtained through numerical simulations that deform the system by imposing a strain. Let us focus in the next section on the possibility to achieve the same qualitative behavior using only equilibrium simulations.

\subsection{Mooney - Rivlin theory detects the transition}

\begin{figure*}[t]
\centering
\includegraphics[width=0.32\textwidth]{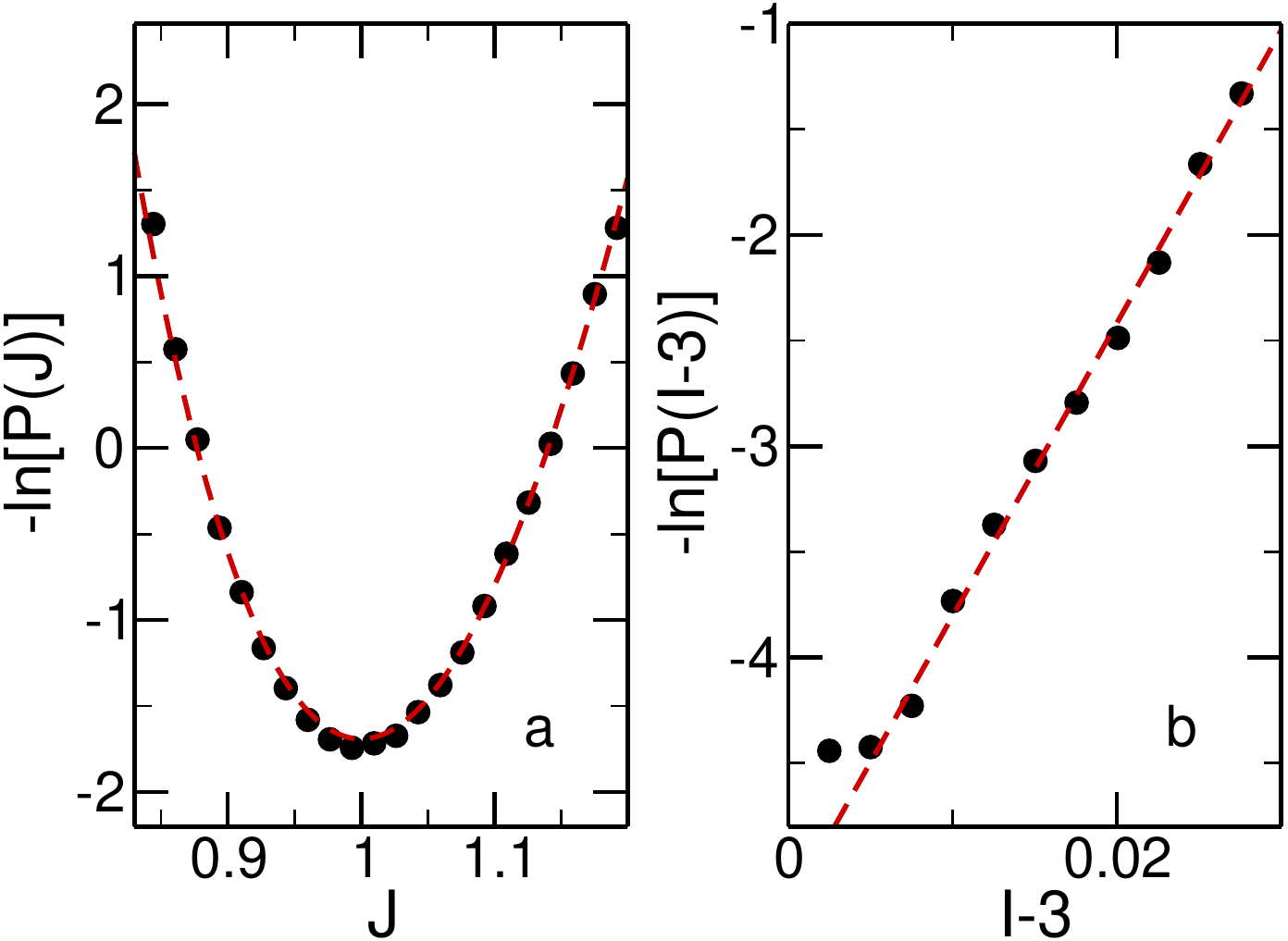}
\includegraphics[width=0.32\textwidth]{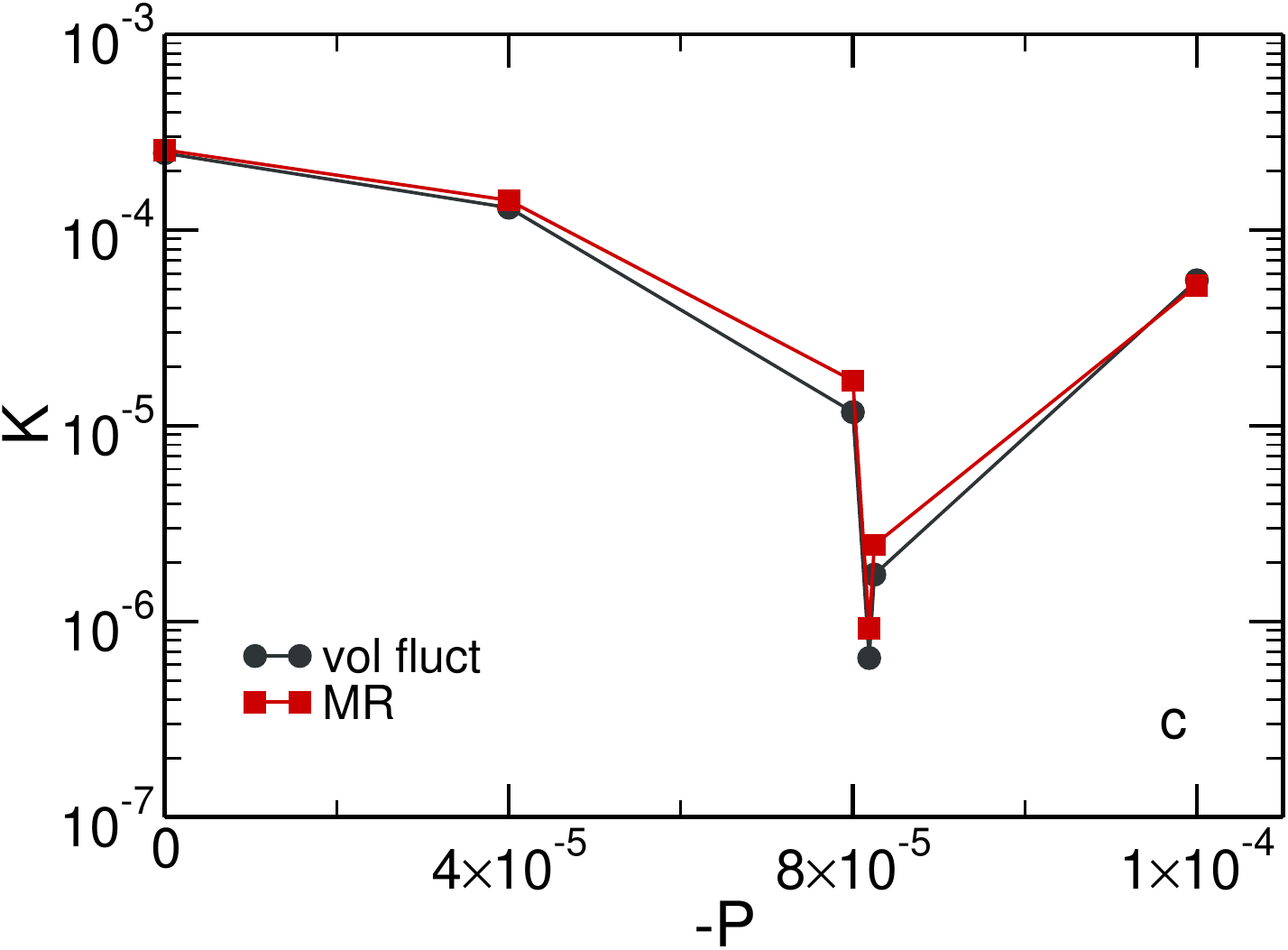}
\includegraphics[width=0.32\textwidth]{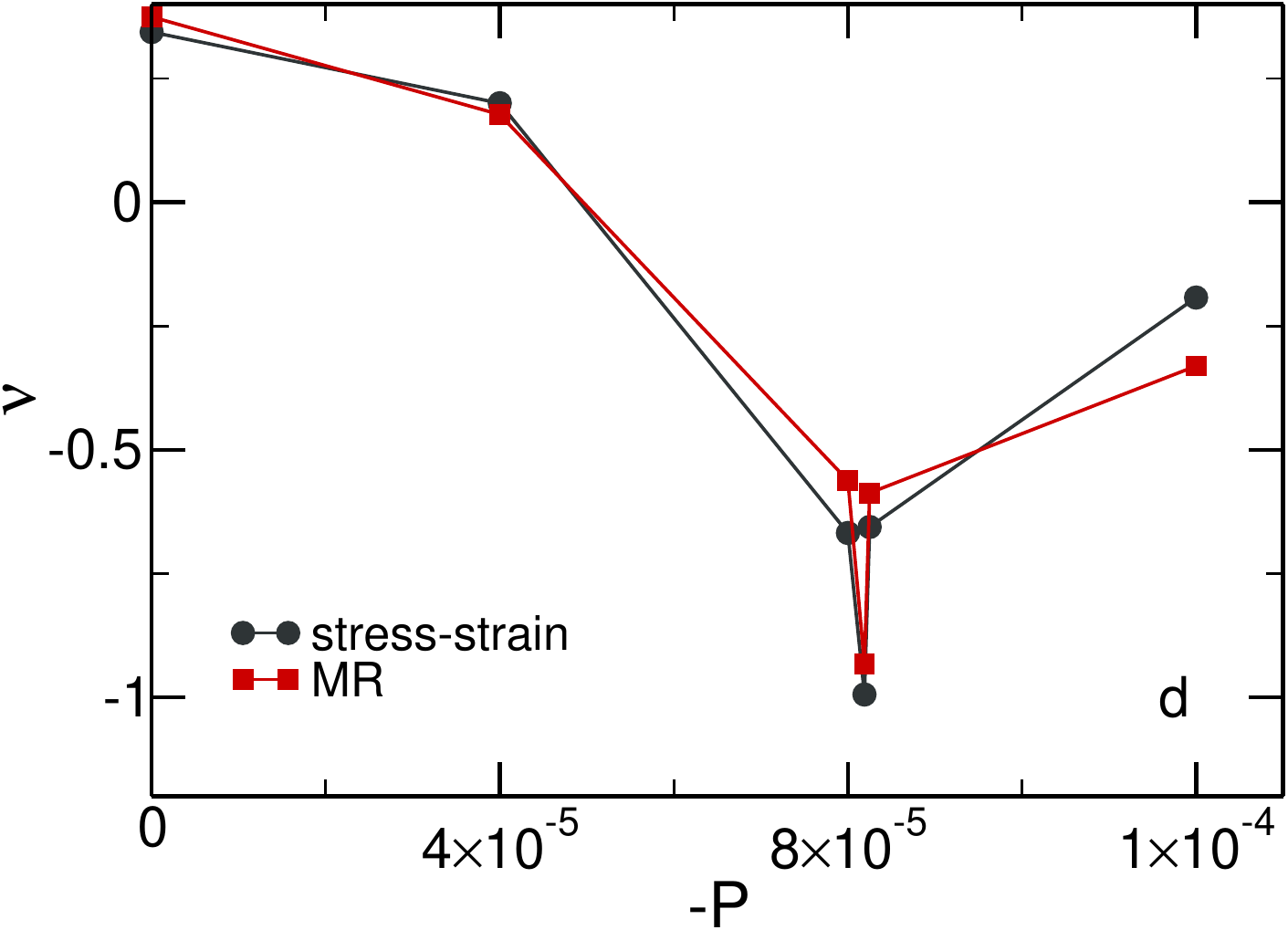}
\caption{(a, b) Symbols show the logarithm of the inverse of the probability distribution of the two strain invariants $J$ (a) and $I$ (b) for the diamond system with $c=0.35\%$ at $P=0$. 
Red dotted curves are fit restrained to values close to zero where the agreement with a $\frac{(J-1)^2}{J}$ and a linear form is reliable; (c) $K$ and (d) $\nu$ as a function of $-P$ for the diamond network with $c=0.35\%$ system, calculated from simulations (circles) and within MR framework (squares).}
\label{fig:MR_comp}
\end{figure*}

We ensure that the elastic properties that we observe in our system are equilibrium ones by computing the elastic moduli across the transition within the framework of the Mooney-Rivlin theory. This relies on phenomenological assumptions and, in the case of bulk systems, it only involves the analysis of fluctuations, in  analogy to what is normally done to evaluate $K$ from volume fluctuations.
In particular, as discussed in Methods,  we obtain strain invariants analysing appropriate fluctuations through Eq.~\eqref{eq:strain} and then, relying on Eq.~\eqref{eq:energyMR}, we fit the distribution of the strain invariants either with a quadratic or with a linear form, as shown in Fig.\ref{fig:MR_comp}(a)(b), respectively for invariants $J$ and $I$  for the diamond system with $c=0.35$\% as a function of pressure.
Doing so, we find a good agreement with the theoretical predictions (Eq.~\eqref{eq:energyMR}), except for minor deviations in Fig.\ref{fig:MR_comp}(b) at small deformations, that can be ascribed to enhanced flexibility close to the unstressed state. 
In Fig.~\ref{fig:MR_comp} (c) and (d)  we then report the values of $K$ and $\nu$ obtained by the MR approach and directly compare them with the results of the simulations, finding very similar results for the moduli in the two cases.  For the bulk modulus, that is estimated from equilibrium fluctuations in both approaches, the agreement is remarkable, providing evidence that the chosen reference state and the phenomenological theory are correct. In the case of the Poisson's ratio, the agreement is slightly less quantitative, but overall very satisfactory and reproducing the minimum  and the occurrence of auxetic behavior in both approaches. The small deviations that are observed can be ascribed to the fact that $\nu$  is obtained indirectly by combining $K$ and $G$, that are directly calculated in the MR approach, via the linear elasticity relation $\nu = \frac{3K-2G}{2(3K+G)}$. As for the other two elastic moduli, $G$ and $Y$ (not shown), we also find comparable values at all pressures. However, data for $Y$ are the ones where larger differences between MR and stress-strain simulations are observed, again because it is calculated indirectly. We also note that $Y$ is also the modulus showing the largest statistical error when estimated stress-strain simulations, particularly close to the minimum where it becomes very small. 
Notwithstanding this, the present results confirm that both methods are quite accurate to calculate the Poisson's ratio of polymer networks and that, for a qualitative assessment, MR can be used without the need to perform time-consuming stress-strain simulations for each state point.

\subsection{The influence of the density on the elastic properties of the network}
\label{sec:density}

As stated previously, one of the key control parameters to enhance fluctuations and get closer to the transition point is the degree of connectivity of the network. In the diamond network, this property is strictly linked to the number density of the system, as $c$ directly determines the (homogeneous) strand length. 
However, the situation is different for disordered networks where we can vary $c$ and $\rho$ independently, by varying the assembly volume. Since, we found that  the system density at $P=0$ is the same for the occurrence of the HAT in both \textit{Diam} and \textit{Diso} networks, although $c$ is different, respectively $\sim 0.35$\% and  $\sim 1$\, in this section we  investigate more accurately the role of the density  in controlling elastic properties.

In particular, we consider four systems: two networks are prepared having a similarly low density but  $c\sim 3\%$ and $7\%$, respectively, and hence called $L_3$ and $L_7$ networks; another two networks are assembled at a high density, similar for the two cases, but  using again the same two different $c$ values. These are referred as $H_3$ and $H_7$, respectively. 
In Table~\ref{tab:HLsys} we gather details about all four systems.

\begin{figure*}[t]
\centering
\includegraphics[width=0.24\textwidth]{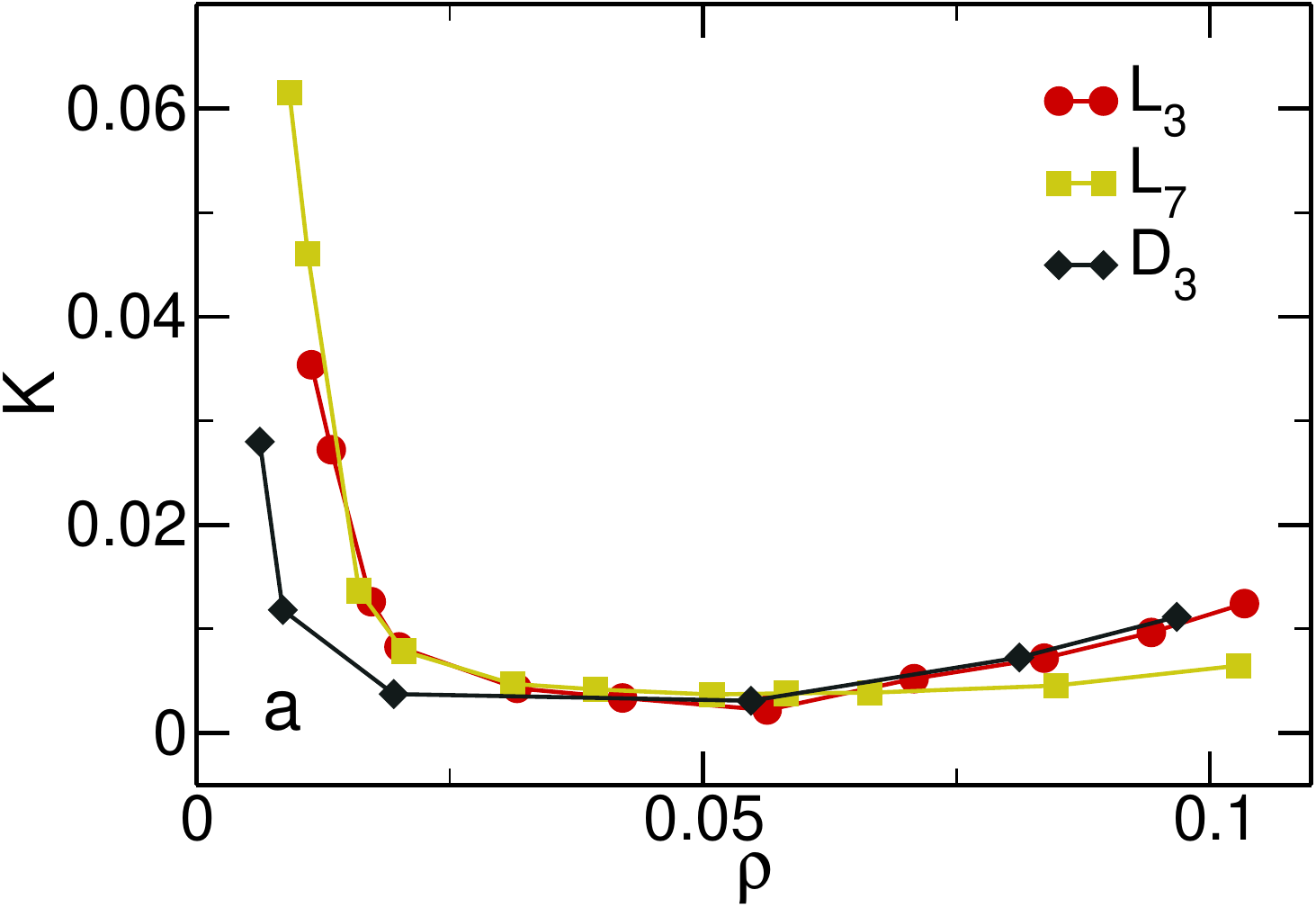}
\includegraphics[width=0.24\textwidth]{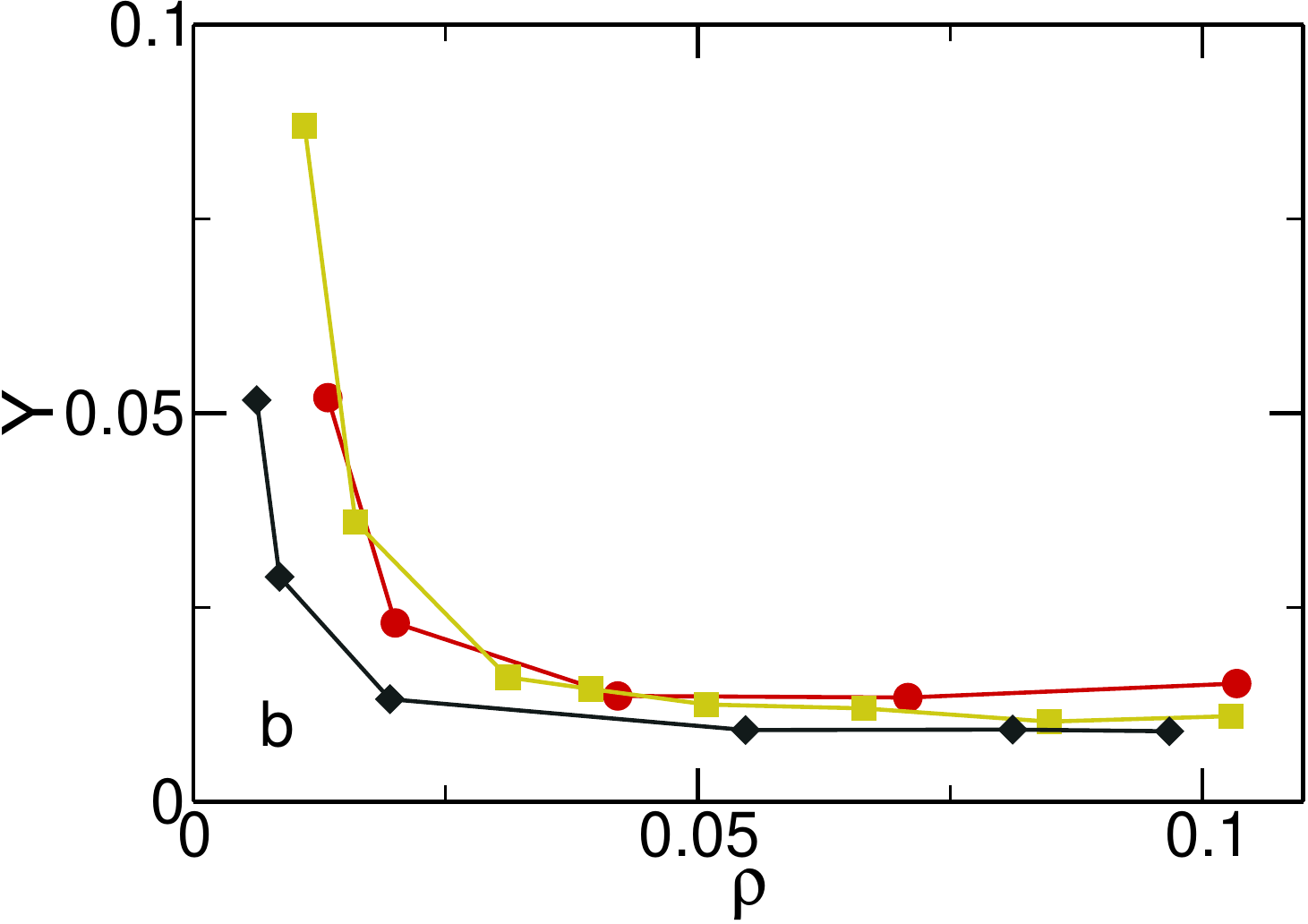}
\includegraphics[width=0.235\textwidth]{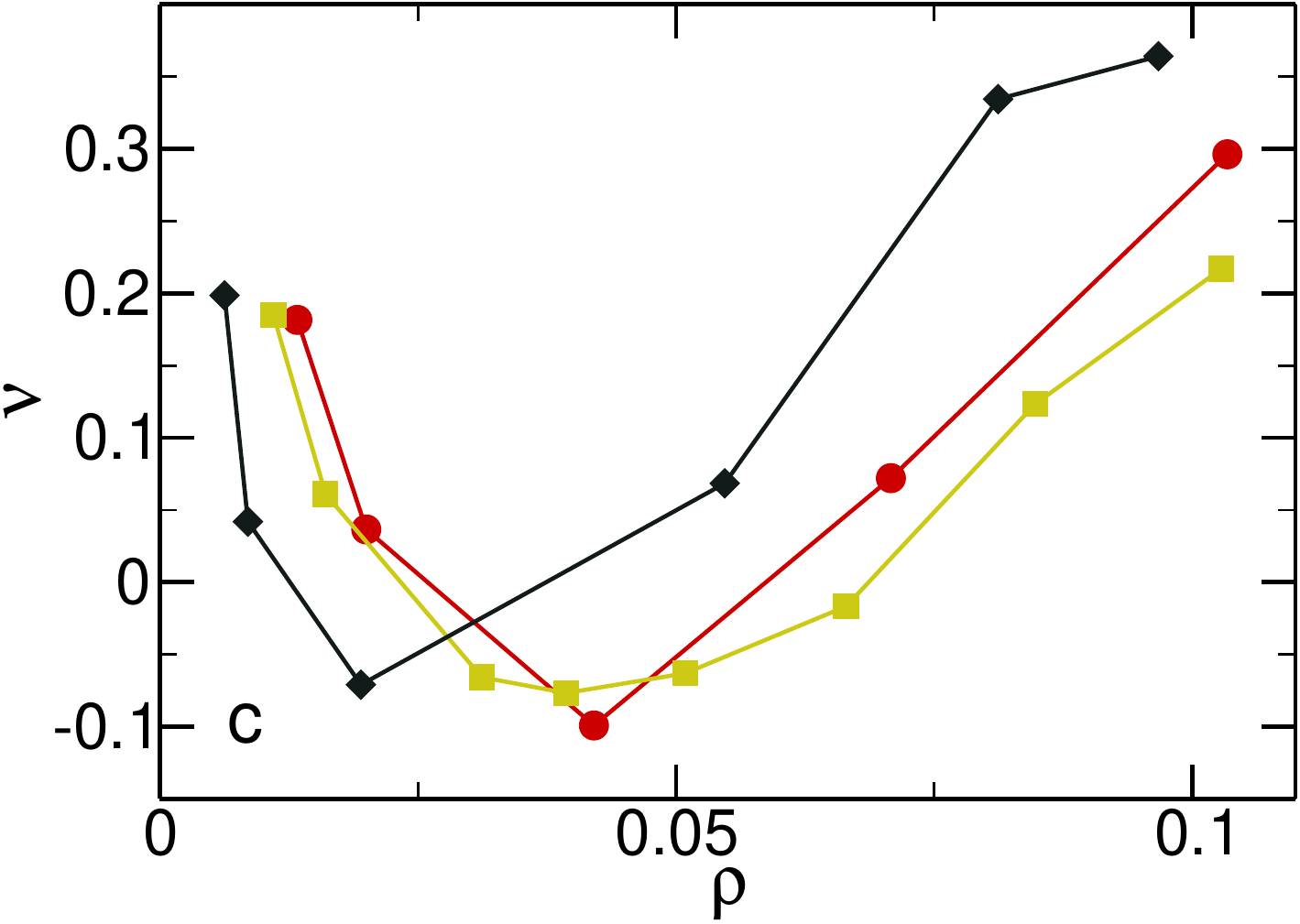}
\includegraphics[width=0.24\textwidth]{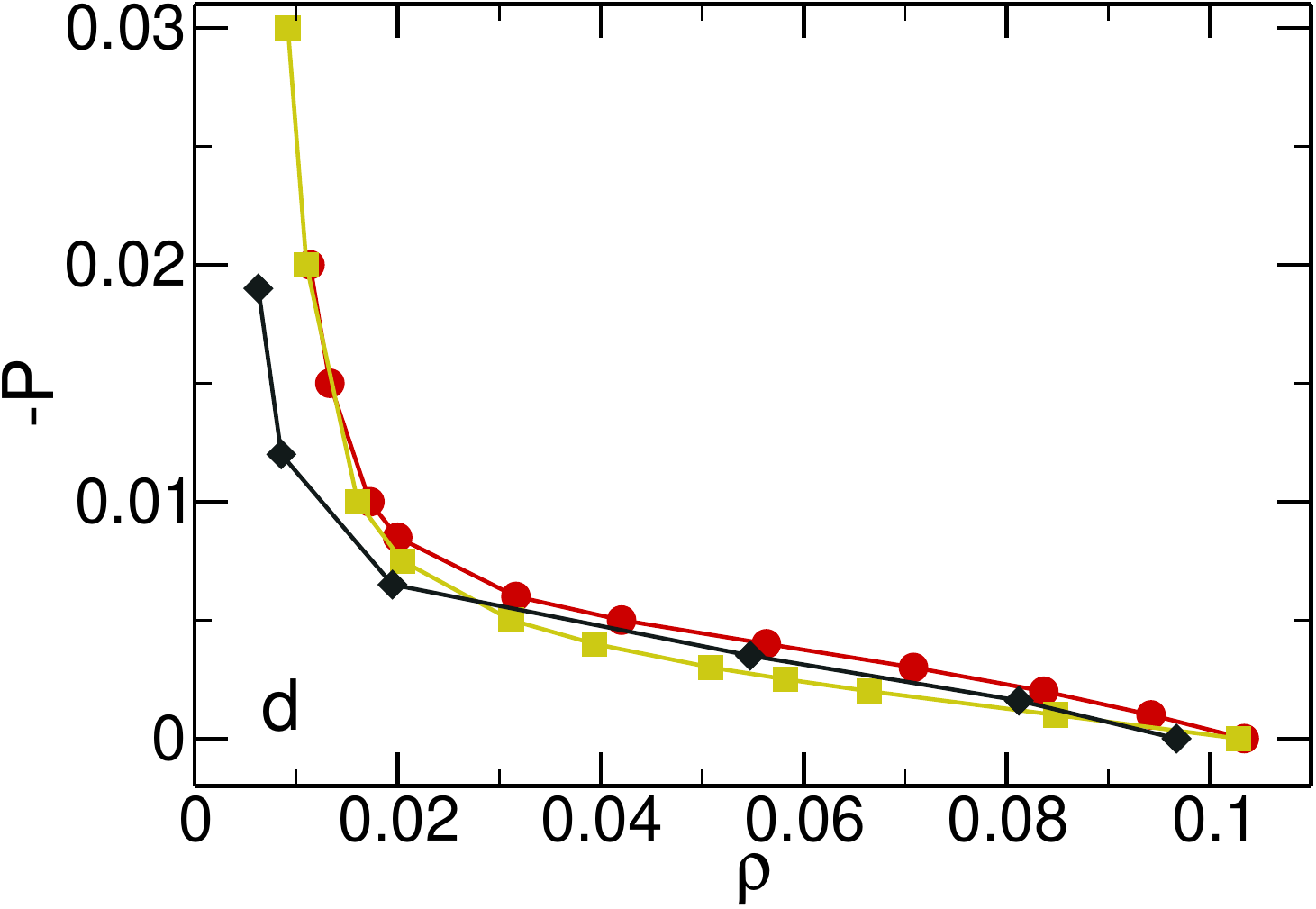}
\includegraphics[width=0.24\textwidth]{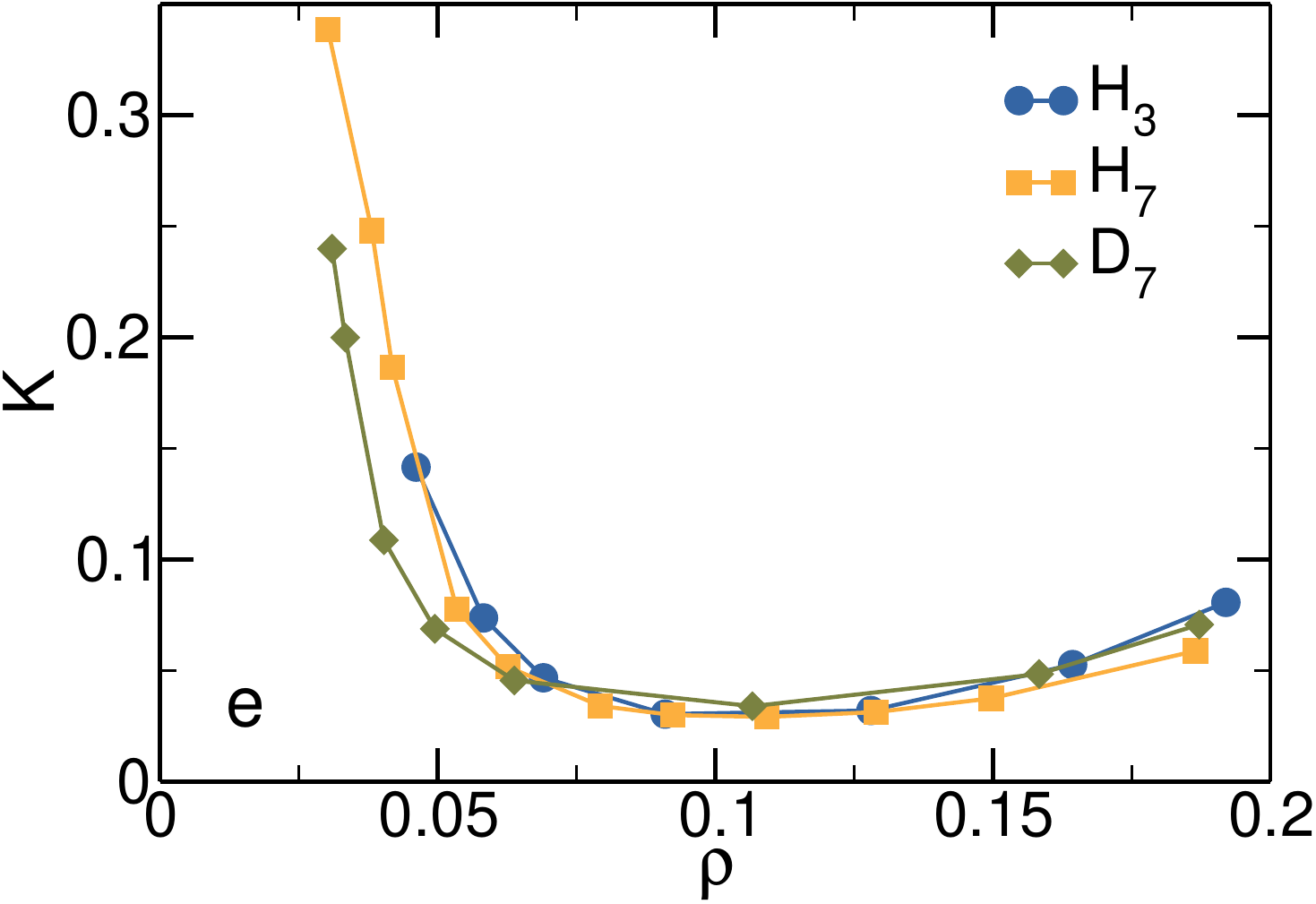}
\includegraphics[width=0.24\textwidth]{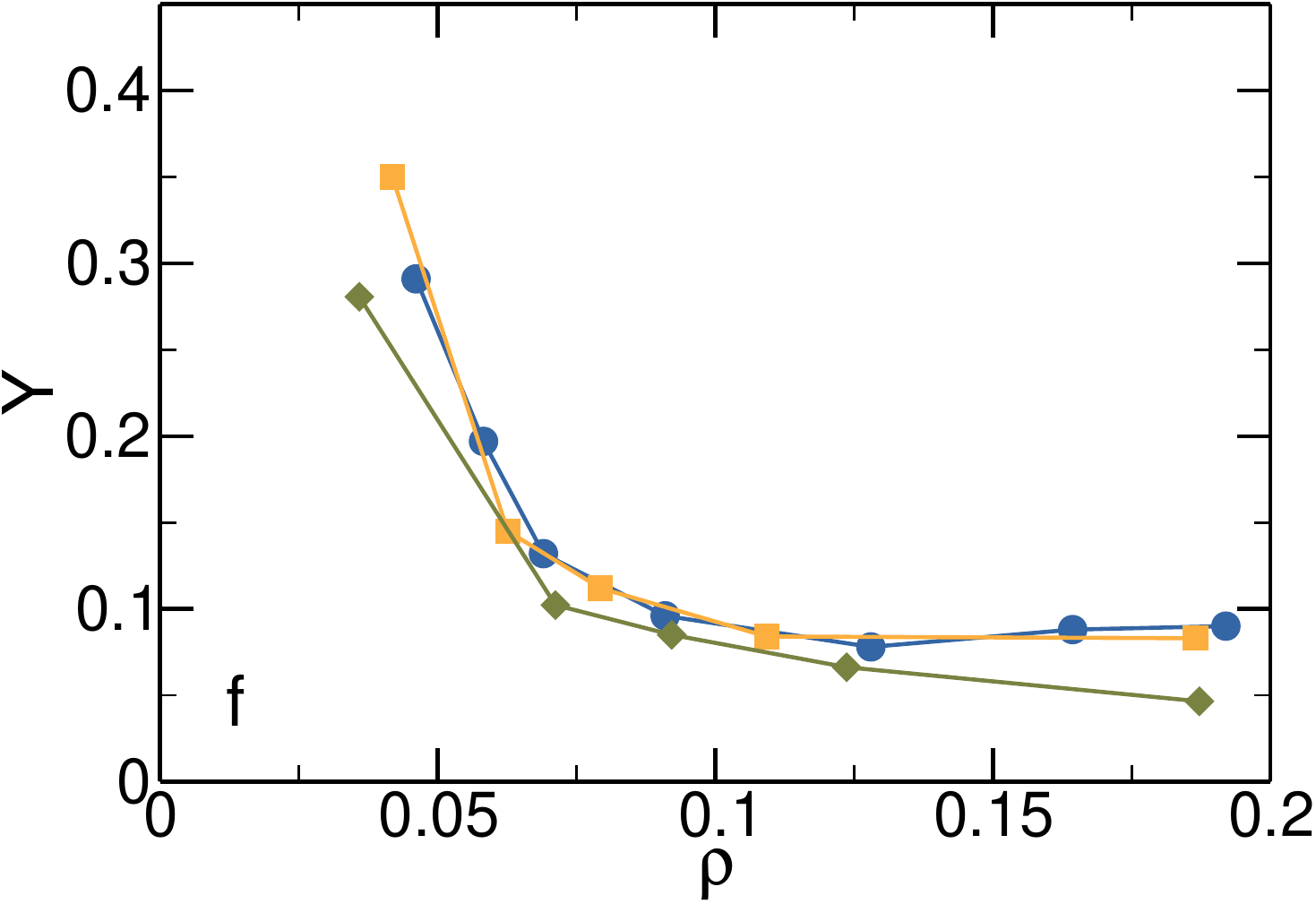}
\includegraphics[width=0.24\textwidth]{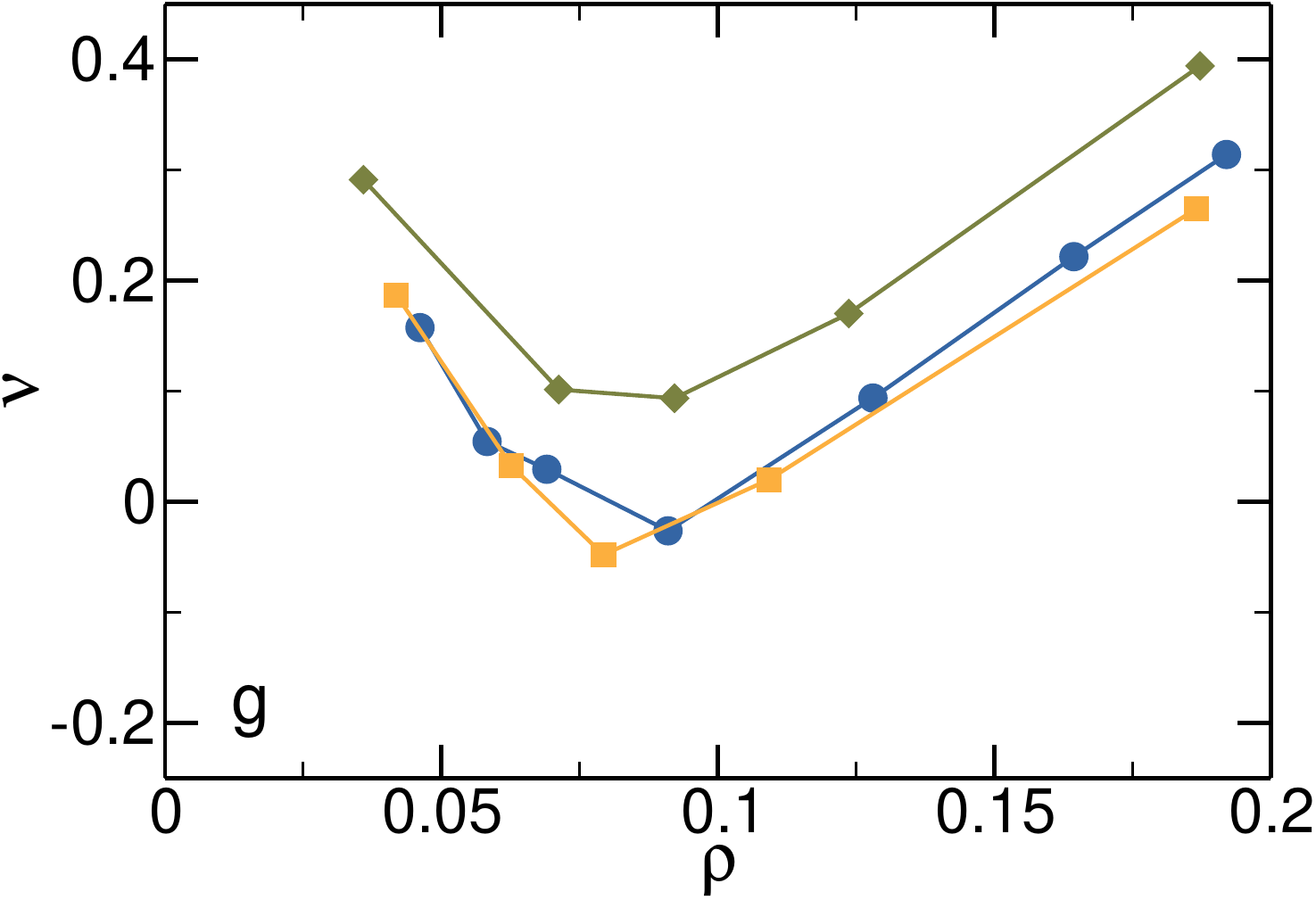}
\includegraphics[width=0.245\textwidth]{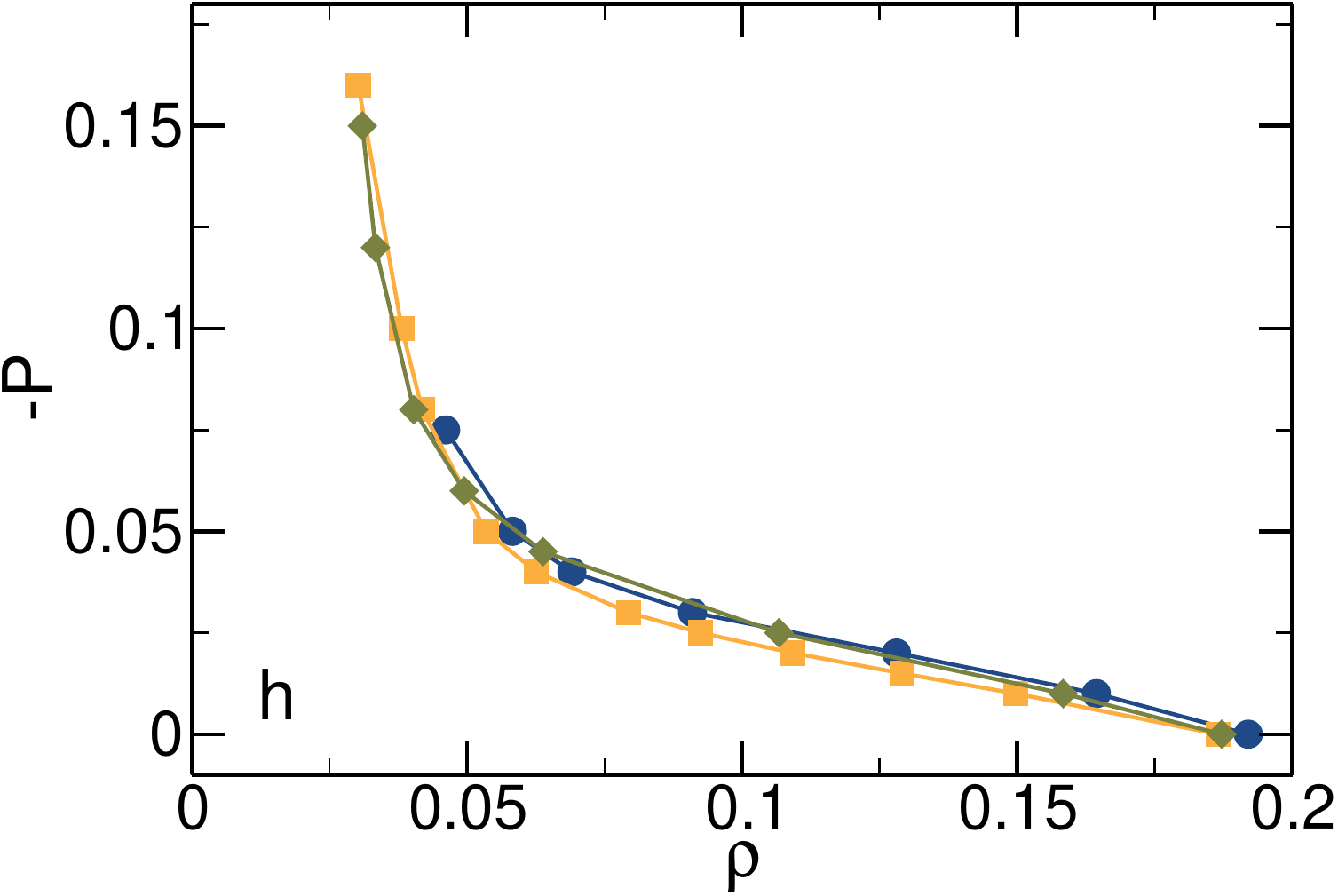}
\caption{Elastic moduli as a function of  $\rho$ for the low-density ($L_3,L_7,D_3$) (top) and the high-density ($H_3,H_7,D_7$) (bottom) networks: $K$ (a),(e), $Y$ (b),(f), and $\nu$ (c),(g), respectively. The corresponding equations of state, $P$ vs $\rho$,  are also reported in (d),(h).}
\label{fig:dens_comp}
\end{figure*}

We investigate elastic properties of the four systems by equilibrium and stress-strain simulations. We are therefore able to obtain all elastic moduli of the different systems and compare among them.
We report the results for $K,Y,\nu$ as a function of $\rho$ for the four systems in Fig.\ref{fig:dens_comp}. 
Interestingly, we find almost superimposed values for all the investigated moduli of the disordered networks having a similar density. In particular, for both $L$ and $H$ systems, the bulk and Young moduli initially undergo a rapid decrease with increasing $\rho$, then go through to a minimum and then increase again. The agreement also holds for the Poisson's ratio. We note that $\nu$ assumes moderate negative values in an ample density interval for the $L$ systems, while it barely becomes negative for $H$ ones.

To see whether this behavior also holds for regular networks, we consider here the corresponding diamond systems having the same $P=0$ densities as well. These are referred as $D_3$ and $D_7$ to be compared with the $L_3,L_7$ and the $H_3,H_7$ pairs, respectively. The corresponding elastic moduli are also reported in Fig.~\ref{fig:dens_comp}, showing a different behavior. First of all, minima are shifted towards smaller densities for all the moduli. Such an effect is small for the bulk modulus as well as for $Y$, but is found to be quite pronounced for the Poisson's ratio, especially for the $D_7$ system. Here, also a much larger value of $\nu$ is found and in general, the values of the moduli are different from those of the disordered networks. Interestingly, while the moduli are not the same, the equation of state of the networks are rather similar, as shown in Fig.~\ref{fig:dens_comp}. This confirms that thermodynamically the system are similar when compared at the same density, but their elastic properties are quantitatively, not qualitatively, different for ordered networks. 

\begin{figure*}[t]
\centering
\includegraphics[width=0.24\textwidth]{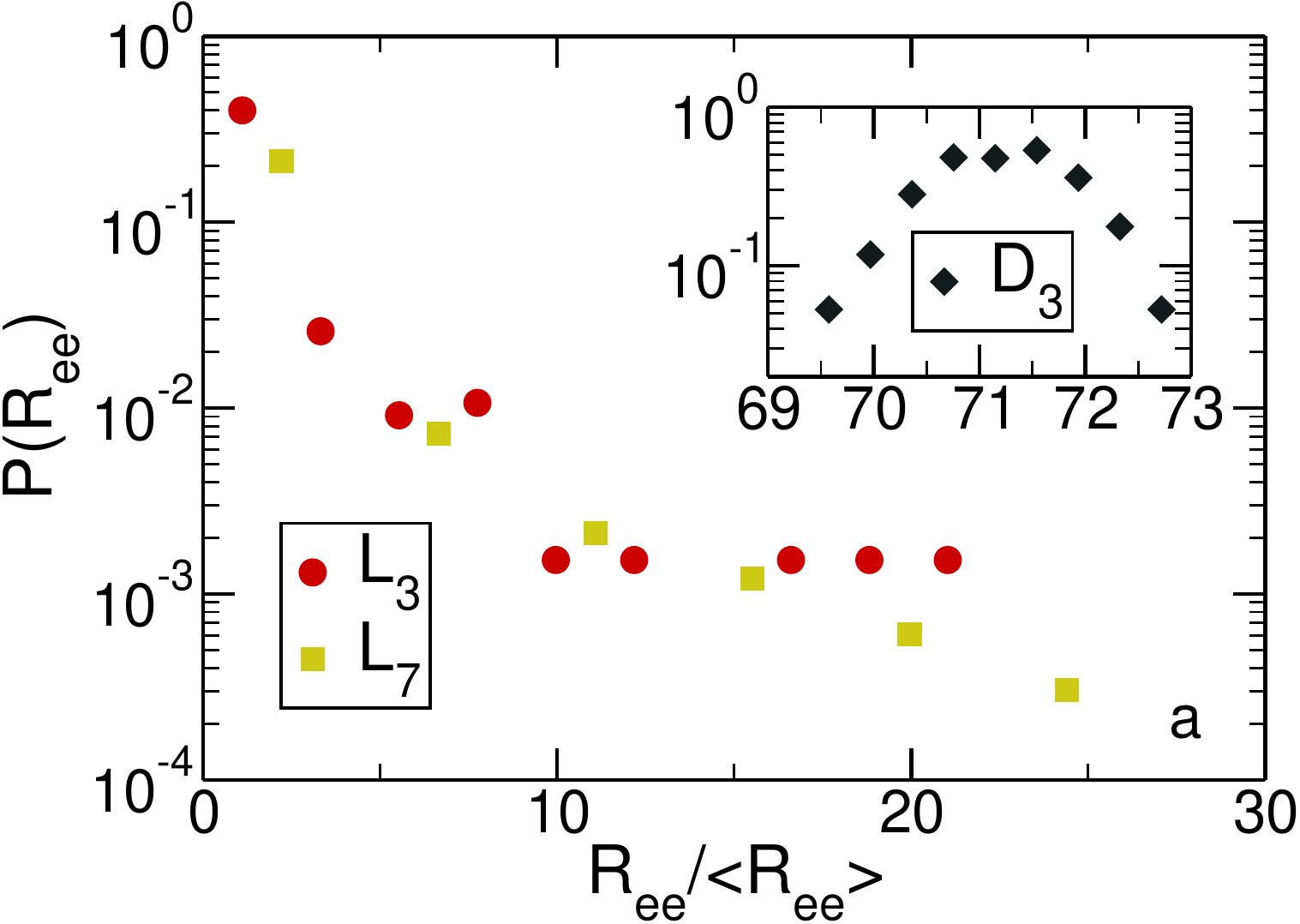}
\includegraphics[width=0.24\textwidth]{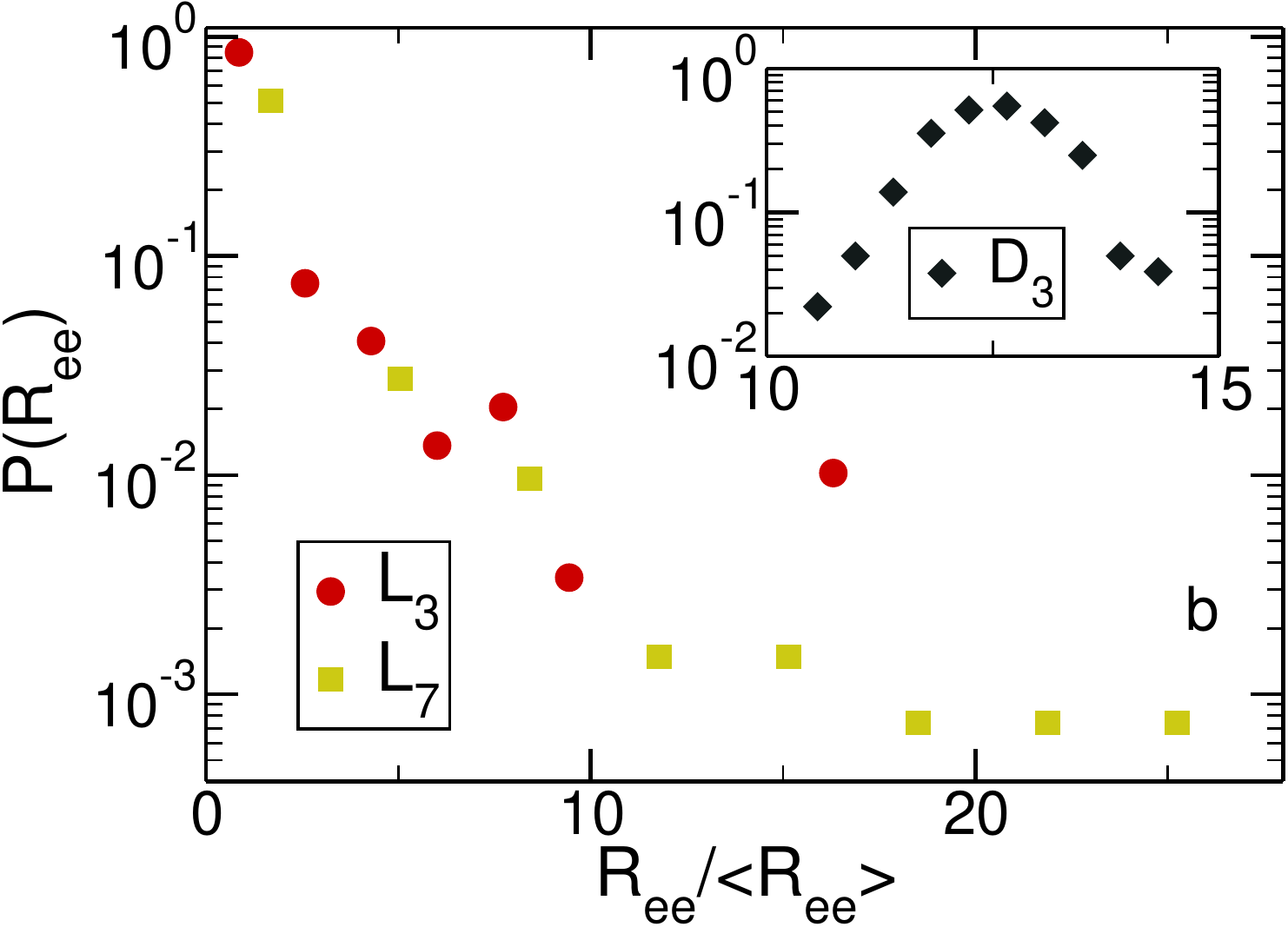}
\includegraphics[width=0.24\textwidth]{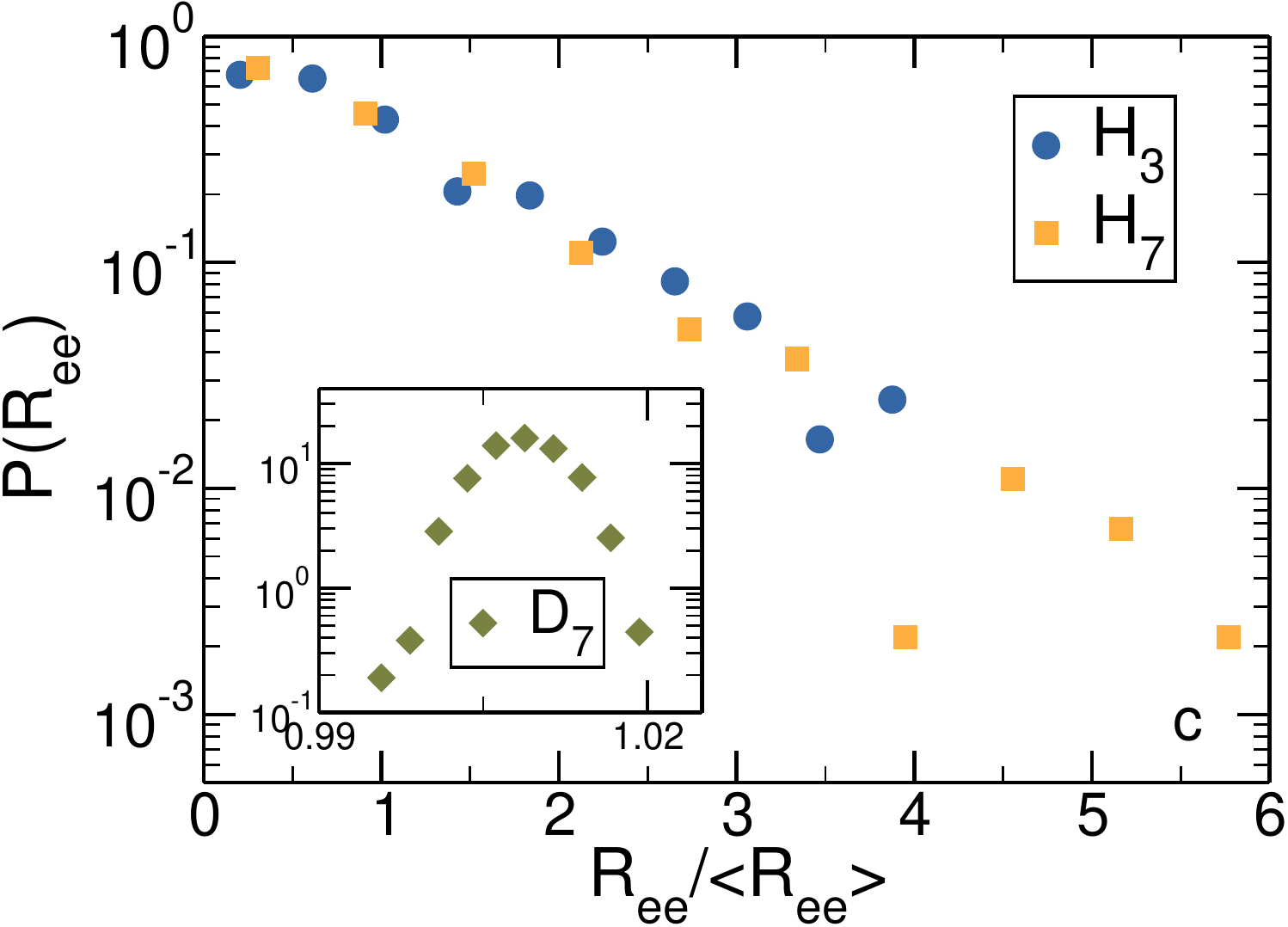}
\includegraphics[width=0.24\textwidth]{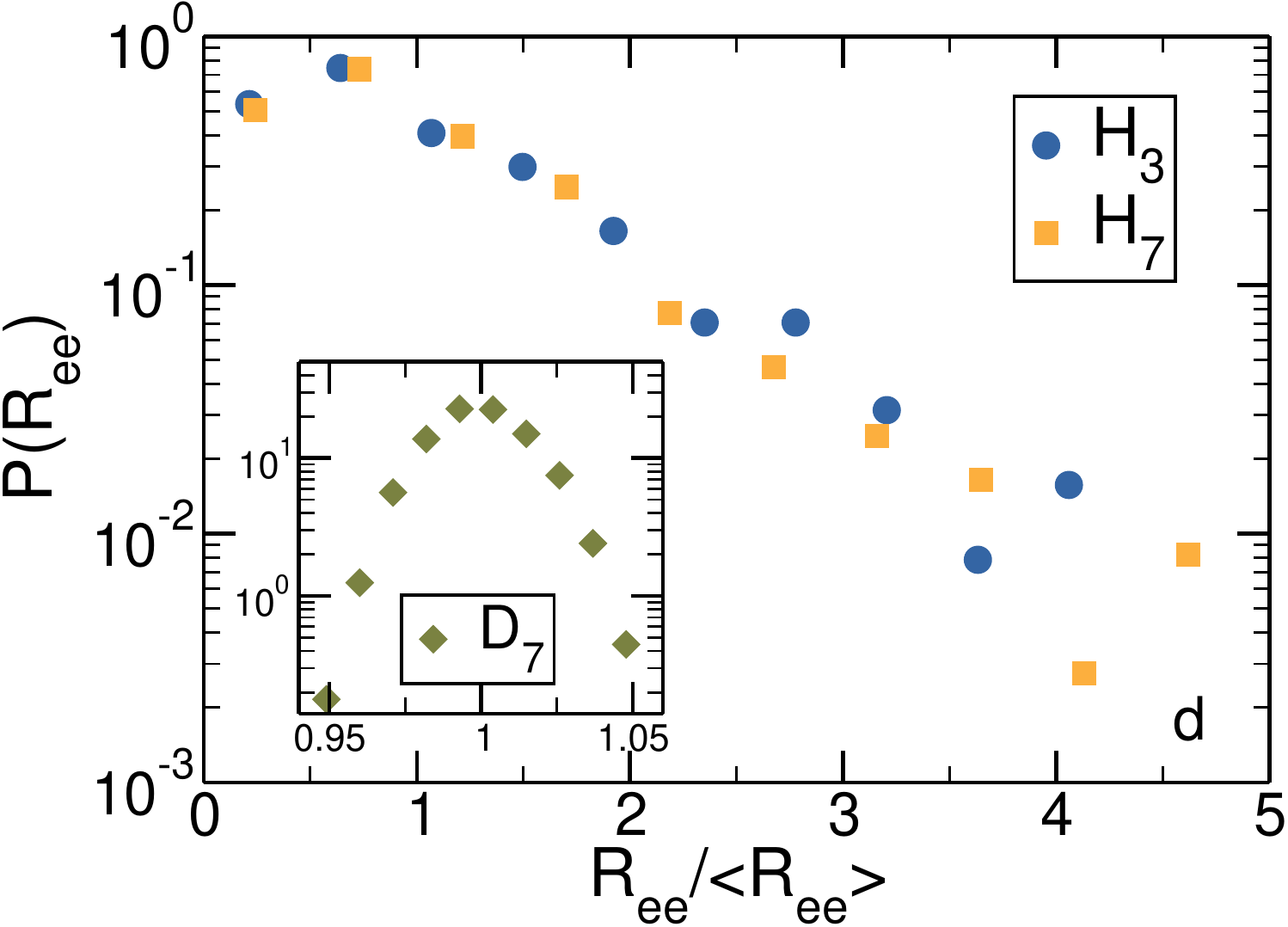}
\caption{Left: End-to-end length distribution for $L_3$,$L_7, D_7$ networks at two representative densities: (a) low ($P=-4.5\times 10^{-2},-5\times 10^{-2}, P=-1.6\times 10^{-2}$) and (b) high ($P=0,0, 0$), respectively. 
Right: The same quantity is reported for the $H_3,H_7,D_7$ systems (c) at ($P=-1\times 10^{-1},-1.1\times 10^{-1},-8\times 10^{-2}$) and (d) at ($P=-2.4\times 10^{-3},-2\times 10^{-4}, 0$), respectively. Results for the diamond systems are shown in the inset.}
\label{fig:Ree_resc}
\end{figure*}

This discrepancy can be rationalized in terms of the topology of the systems. Disordered and ordered networks have indeed  extremely different distributions of the chain lengths. Namely, while in ordered systems the chain lengths are fixed by $c$, for disordered networks they follow an exponential function with a $c$-dependent base. This structural difference clearly influences the elastic behavior.~\cite{Sorichetti2021} To quantify this effect, we have also calculated end-to-end lengths of each chain, $R_{ee}$,~\cite{Rubinstein2014} and the corresponding distributions $P(R_{ee})$ for a representative low and a representative high density case, for all the systems taken into account. The corresponding results are shown in Fig.~\ref{fig:Ree_resc}. We find that the $R_{ee}$ distribution for the disordered systems follows an exponential decay, which again only depends on the density $\rho(P=0)$, since the two pairs of systems $H_3,H_7$ and $L_3,L_7$ have very similar distributions both in the high-$\rho$ and in the low-$\rho$ regime. Note that this comparison is robust, as indicated by the presence of a maximum for the high-density case in both $H_3,H_7$ systems. The fact that these distributions show such equivalence is, however, at odds with the behavior of the distributions of the chemical length of strands, whose exponential decay has a $c$ dependence of the slope. Conversely, the situation is very different for the diamond system, for which $R_{ee}$ shows a Gaussian distribution that is strongly peaked close to the mean value of the distribution, due to the uniform chain length of the network.



\section{Discussion and Conclusions}

In this paper, we have investigated the recently reported auxetic behavior of polymer networks at low connectivity under tension by means of extensive numerical simulations, both equilibrium and stress-strain, as well as through the framework of the phenomenological Mooney-Rivlin theory of elasticity.

We found that the hyper-auxetic behavior, corresponding to $\nu=-1$, is linked to the emergence of critical-like phenomena, arising close to the mechanical instability. While in our previous work~\cite{Ninarello2022}, we only found evidence of this phenomenon in ordered networks, here we confirm that it also takes place in disordered systems. In particular, we found quantitative behavior of the elastic moduli between ordered and disordered systems. This suggests that the HAT is a generic phenomenon of low-connectivity polymer networks under tension, independently of the system geometry. Although the agreement holds also for the probability distribution of the order parameter, within the present numerical accuracy, we cannot assert whether the transition follows the Ising universality class or not. Moreover, being disordered networks the ones that are close to realistic topologies, that can be realized in experiments, we believe that the present findings can stimulate future experimental work to confirm these intriguing predictions. 

The present numerical results have been confirmed by elastic properties calculations in the context of a phenomenological theory of linear elasticity of solids, the Mooney-Rivlin theory. We found that the theoretical results, based on equilibrium simulations are coherent with those obtained by simulations,  also when strain is applied, once more confirming the occurrence of the HAT. 
As a consequence, we also deduce that our system is in an elastic regime throughout the HAT, as we expected, since the MR theory lay its foundations in response of solids at small deformations, far from the plastic regime.
This provides another indication of the reliability of our findings for the hyper-auxetic behavior of polymer networks.

We also investigated the effect of the network density on its elastic properties. In particular, we found that the density at zero pressure is the dominant control parameter which determines the moduli of disordered polymer networks, independently of their crosslinker concentration. Instead, for ordered polymer networks, where density and crosslinker concentration are coupled, such a statement does not hold. Indeed, the moduli of the diamond networks are qualitatively similar, but quantitatively different, in density, due to the underlying homogeneous strand length distribution. This points to the dominant role of short strands in determining the elastic properties of the network.

The importance of chain polydispersity in determining the elasticity of the system was reported in Ref.~\cite{Sorichetti2021} as well. In particular, under tension, we found that short chains immediately respond by stretching, while long chains have more room to rearrange. The results reported in this paper, on the one hand, confirm these previous findings and, on the other hand, allow a comparison between ordered and disordered systems. Furthermore, we observe that disordered systems with similar elastic behavior feature a comparable chain extension distribution marking again the importance of the features of the chain population. Our present results on end-to-end length distribution thus suggest as well that the elastic behavior of the disordered networks is strongly determined by $\rho(P=0)$ , appearing to be only indirectly influenced by the crosslinker concentration. Yet, this reasoning does not hold for the diamond network due to their intrinsically different topology.

Notably, the results for disordered networks reported in this work not only deepen our understanding of this phenomenon, but also clarify the primary role of thermodynamics in this transition. Indeed, we find confirmation that an intimate link between a mechanical and a thermodynamical instability exists in these systems. It will be important to fully understand this connection in the future, particularly in relation to the universality class of the transition, which does not fully obey, within the current numerical resolution, the Ising behavior for the analog liquid-gas separation. In addition, the order parameter is found to be slightly different, amounting to a coupling of the density with non-bonded energy only, highlighting the important role of entropy in the HAT phenomenology. To this aim, it also remains to understand what will happen in the presence of inter-particle attraction, e.g. by changing temperature in responsive polymer hydrogels.  Finally, it is still unclear how to reconcile the present observations at negative pressures and low temperature with  the so-called Volume Phase Transition observed in thermoresponsive polymer networks.  Future work will be focused to unveil how the temperature affects the occurrence of the HAT.



\section{Acknowledgments}
EZ acknowledges financial support from ICSC -- Centro Nazionale di Ricerca in High Performance Computing, Big Data and Quantum Computing, funded by European Union -- NextGenerationEU - PNRR, Missione 4 Componente 2 Investimento 1.4.
The authors gratefully acknowledge CINECA ISCRA for HPC resources. We thank Prof. L Rovigatti for pointing us to the complete version of Eq. \ref{eq:energyMR}.
\section{Data Availability}
The data that support the findings of this study are available at \url{https://doi.org/10.24435/materialscloud:64-wr} and from the corresponding author upon reasonable request.
\bibliography{bibliography.bib}

\begin{thebibliography}{48}%
\makeatletter
\providecommand \@ifxundefined [1]{%
 \@ifx{#1\undefined}
}%
\providecommand \@ifnum [1]{%
 \ifnum #1\expandafter \@firstoftwo
 \else \expandafter \@secondoftwo
 \fi
}%
\providecommand \@ifx [1]{%
 \ifx #1\expandafter \@firstoftwo
 \else \expandafter \@secondoftwo
 \fi
}%
\providecommand \natexlab [1]{#1}%
\providecommand \enquote  [1]{``#1''}%
\providecommand \bibnamefont  [1]{#1}%
\providecommand \bibfnamefont [1]{#1}%
\providecommand \citenamefont [1]{#1}%
\providecommand \href@noop [0]{\@secondoftwo}%
\providecommand \href [0]{\begingroup \@sanitize@url \@href}%
\providecommand \@href[1]{\@@startlink{#1}\@@href}%
\providecommand \@@href[1]{\endgroup#1\@@endlink}%
\providecommand \@sanitize@url [0]{\catcode `\\12\catcode `\$12\catcode
  `\&12\catcode `\#12\catcode `\^12\catcode `\_12\catcode `\%12\relax}%
\providecommand \@@startlink[1]{}%
\providecommand \@@endlink[0]{}%
\providecommand \url  [0]{\begingroup\@sanitize@url \@url }%
\providecommand \@url [1]{\endgroup\@href {#1}{\urlprefix }}%
\providecommand \urlprefix  [0]{URL }%
\providecommand \Eprint [0]{\href }%
\providecommand \doibase [0]{http://dx.doi.org/}%
\providecommand \selectlanguage [0]{\@gobble}%
\providecommand \bibinfo  [0]{\@secondoftwo}%
\providecommand \bibfield  [0]{\@secondoftwo}%
\providecommand \translation [1]{[#1]}%
\providecommand \BibitemOpen [0]{}%
\providecommand \bibitemStop [0]{}%
\providecommand \bibitemNoStop [0]{.\EOS\space}%
\providecommand \EOS [0]{\spacefactor3000\relax}%
\providecommand \BibitemShut  [1]{\csname bibitem#1\endcsname}%
\let\auto@bib@innerbib\@empty
\bibitem [{\citenamefont {Lakes}(1987)}]{Lakes1987}%
  \BibitemOpen
  \bibfield  {author} {\bibinfo {author} {\bibfnamefont {R.}~\bibnamefont
  {Lakes}},\ }\href@noop {} {\bibfield  {journal} {\bibinfo  {journal}
  {Science}\ }\textbf {\bibinfo {volume} {235}},\ \bibinfo {pages} {1038}
  (\bibinfo {year} {1987})}\BibitemShut {NoStop}%
\bibitem [{\citenamefont {Evans}\ \emph {et~al.}(1991)\citenamefont {Evans},
  \citenamefont {Nkansah}, \citenamefont {Hutchinson},\ and\ \citenamefont
  {Rogers}}]{evans1991molecular}%
  \BibitemOpen
  \bibfield  {author} {\bibinfo {author} {\bibfnamefont {K.~E.}\ \bibnamefont
  {Evans}}, \bibinfo {author} {\bibfnamefont {M.}~\bibnamefont {Nkansah}},
  \bibinfo {author} {\bibfnamefont {I.}~\bibnamefont {Hutchinson}}, \ and\
  \bibinfo {author} {\bibfnamefont {S.}~\bibnamefont {Rogers}},\ }\href@noop {}
  {\bibfield  {journal} {\bibinfo  {journal} {Nature}\ }\textbf {\bibinfo
  {volume} {353}},\ \bibinfo {pages} {124} (\bibinfo {year}
  {1991})}\BibitemShut {NoStop}%
\bibitem [{\citenamefont {Greaves}\ \emph {et~al.}(2011)\citenamefont
  {Greaves}, \citenamefont {Greer}, \citenamefont {Lakes},\ and\ \citenamefont
  {Rouxel}}]{greaves2011poisson}%
  \BibitemOpen
  \bibfield  {author} {\bibinfo {author} {\bibfnamefont {G.~N.}\ \bibnamefont
  {Greaves}}, \bibinfo {author} {\bibfnamefont {A.}~\bibnamefont {Greer}},
  \bibinfo {author} {\bibfnamefont {R.~S.}\ \bibnamefont {Lakes}}, \ and\
  \bibinfo {author} {\bibfnamefont {T.}~\bibnamefont {Rouxel}},\ }\href@noop {}
  {\bibfield  {journal} {\bibinfo  {journal} {Nature materials}\ }\textbf
  {\bibinfo {volume} {10}},\ \bibinfo {pages} {823} (\bibinfo {year}
  {2011})}\BibitemShut {NoStop}%
\bibitem [{\citenamefont {Bose}, \citenamefont {Roy},\ and\ \citenamefont
  {Bandyopadhyay}(2012)}]{Bose2012}%
  \BibitemOpen
  \bibfield  {author} {\bibinfo {author} {\bibfnamefont {S.}~\bibnamefont
  {Bose}}, \bibinfo {author} {\bibfnamefont {M.}~\bibnamefont {Roy}}, \ and\
  \bibinfo {author} {\bibfnamefont {A.}~\bibnamefont {Bandyopadhyay}},\ }\href
  {\doibase 10.1016/j.tibtech.2012.07.005} {\bibfield  {journal} {\bibinfo
  {journal} {Trends in Biotechnology}\ }\textbf {\bibinfo {volume} {30}},\
  \bibinfo {pages} {546} (\bibinfo {year} {2012})}\BibitemShut {NoStop}%
\bibitem [{\citenamefont {Duncan}\ \emph {et~al.}(2018)\citenamefont {Duncan},
  \citenamefont {Shepherd}, \citenamefont {Moroney}, \citenamefont {Foster},
  \citenamefont {Venkatraman}, \citenamefont {Winwood}, \citenamefont {Allen},\
  and\ \citenamefont {Alderson}}]{Duncan2018}%
  \BibitemOpen
  \bibfield  {author} {\bibinfo {author} {\bibfnamefont {O.}~\bibnamefont
  {Duncan}}, \bibinfo {author} {\bibfnamefont {T.}~\bibnamefont {Shepherd}},
  \bibinfo {author} {\bibfnamefont {C.}~\bibnamefont {Moroney}}, \bibinfo
  {author} {\bibfnamefont {L.}~\bibnamefont {Foster}}, \bibinfo {author}
  {\bibfnamefont {P.}~\bibnamefont {Venkatraman}}, \bibinfo {author}
  {\bibfnamefont {K.}~\bibnamefont {Winwood}}, \bibinfo {author} {\bibfnamefont
  {T.}~\bibnamefont {Allen}}, \ and\ \bibinfo {author} {\bibfnamefont
  {A.}~\bibnamefont {Alderson}},\ }\href {\doibase 10.3390/app8060941}
  {\bibfield  {journal} {\bibinfo  {journal} {Applied Sciences}\ }\textbf
  {\bibinfo {volume} {8}},\ \bibinfo {pages} {941} (\bibinfo {year}
  {2018})}\BibitemShut {NoStop}%
\bibitem [{\citenamefont {Tahir}, \citenamefont {Zhang},\ and\ \citenamefont
  {Hu}(2022)}]{Tahir2022}%
  \BibitemOpen
  \bibfield  {author} {\bibinfo {author} {\bibfnamefont {D.}~\bibnamefont
  {Tahir}}, \bibinfo {author} {\bibfnamefont {M.}~\bibnamefont {Zhang}}, \ and\
  \bibinfo {author} {\bibfnamefont {H.}~\bibnamefont {Hu}},\ }\href {\doibase
  10.1002/pssb.202200324} {\bibfield  {journal} {\bibinfo  {journal} {physica
  status solidi (b)}\ }\textbf {\bibinfo {volume} {259}},\ \bibinfo {pages}
  {2200324} (\bibinfo {year} {2022})}\BibitemShut {NoStop}%
\bibitem [{\citenamefont {Caddock}\ and\ \citenamefont
  {Evans}(1989)}]{Caddock1989}%
  \BibitemOpen
  \bibfield  {author} {\bibinfo {author} {\bibfnamefont {B.~D.}\ \bibnamefont
  {Caddock}}\ and\ \bibinfo {author} {\bibfnamefont {K.~E.}\ \bibnamefont
  {Evans}},\ }\href@noop {} {\bibfield  {journal} {\bibinfo  {journal} {Journal
  of Physics D: Applied Physics}\ }\textbf {\bibinfo {volume} {22}},\ \bibinfo
  {pages} {1877} (\bibinfo {year} {1989})}\BibitemShut {NoStop}%
\bibitem [{\citenamefont {Evans}\ and\ \citenamefont
  {Caddock}(1989)}]{Evans1989}%
  \BibitemOpen
  \bibfield  {author} {\bibinfo {author} {\bibfnamefont {K.~E.}\ \bibnamefont
  {Evans}}\ and\ \bibinfo {author} {\bibfnamefont {B.~D.}\ \bibnamefont
  {Caddock}},\ }\href@noop {} {\bibfield  {journal} {\bibinfo  {journal}
  {Journal of Physics D: Applied Physics}\ }\textbf {\bibinfo {volume} {22}},\
  \bibinfo {pages} {1883} (\bibinfo {year} {1989})}\BibitemShut {NoStop}%
\bibitem [{\citenamefont {Hu}, \citenamefont {Wang},\ and\ \citenamefont
  {Liu}(2011)}]{HongHu2011}%
  \BibitemOpen
  \bibfield  {author} {\bibinfo {author} {\bibfnamefont {H.}~\bibnamefont
  {Hu}}, \bibinfo {author} {\bibfnamefont {Z.}~\bibnamefont {Wang}}, \ and\
  \bibinfo {author} {\bibfnamefont {S.}~\bibnamefont {Liu}},\ }\href@noop {}
  {\bibfield  {journal} {\bibinfo  {journal} {Textile Research Journal}\
  }\textbf {\bibinfo {volume} {81}},\ \bibinfo {pages} {1493} (\bibinfo {year}
  {2011})}\BibitemShut {NoStop}%
\bibitem [{\citenamefont {Gatt}\ \emph {et~al.}(2015)\citenamefont {Gatt},
  \citenamefont {Wood}, \citenamefont {Gatt}, \citenamefont {Zarb},
  \citenamefont {Formosa}, \citenamefont {Azzopardi}, \citenamefont {Casha},
  \citenamefont {Agius}, \citenamefont {Schembri-Wismayer}, \citenamefont
  {Attard}, \citenamefont {Chockalingam},\ and\ \citenamefont
  {Grima}}]{Gatt2015}%
  \BibitemOpen
  \bibfield  {author} {\bibinfo {author} {\bibfnamefont {R.}~\bibnamefont
  {Gatt}}, \bibinfo {author} {\bibfnamefont {M.~V.}\ \bibnamefont {Wood}},
  \bibinfo {author} {\bibfnamefont {A.}~\bibnamefont {Gatt}}, \bibinfo {author}
  {\bibfnamefont {F.}~\bibnamefont {Zarb}}, \bibinfo {author} {\bibfnamefont
  {C.}~\bibnamefont {Formosa}}, \bibinfo {author} {\bibfnamefont {K.~M.}\
  \bibnamefont {Azzopardi}}, \bibinfo {author} {\bibfnamefont {A.}~\bibnamefont
  {Casha}}, \bibinfo {author} {\bibfnamefont {T.~P.}\ \bibnamefont {Agius}},
  \bibinfo {author} {\bibfnamefont {P.}~\bibnamefont {Schembri-Wismayer}},
  \bibinfo {author} {\bibfnamefont {L.}~\bibnamefont {Attard}}, \bibinfo
  {author} {\bibfnamefont {N.}~\bibnamefont {Chockalingam}}, \ and\ \bibinfo
  {author} {\bibfnamefont {J.~N.}\ \bibnamefont {Grima}},\ }\href@noop {}
  {\bibfield  {journal} {\bibinfo  {journal} {Acta Biomaterialia}\ }\textbf
  {\bibinfo {volume} {24}},\ \bibinfo {pages} {201} (\bibinfo {year}
  {2015})}\BibitemShut {NoStop}%
\bibitem [{\citenamefont {Bertoldi}\ \emph {et~al.}(2017)\citenamefont
  {Bertoldi}, \citenamefont {Vitelli}, \citenamefont {Christensen},\ and\
  \citenamefont {van Hecke}}]{Bertoldi2017}%
  \BibitemOpen
  \bibfield  {author} {\bibinfo {author} {\bibfnamefont {K.}~\bibnamefont
  {Bertoldi}}, \bibinfo {author} {\bibfnamefont {V.}~\bibnamefont {Vitelli}},
  \bibinfo {author} {\bibfnamefont {J.}~\bibnamefont {Christensen}}, \ and\
  \bibinfo {author} {\bibfnamefont {M.}~\bibnamefont {van Hecke}},\ }\href
  {\doibase 10.1038/natrevmats.2017.66} {\bibfield  {journal} {\bibinfo
  {journal} {Nature Reviews Materials}\ }\textbf {\bibinfo {volume} {2}}
  (\bibinfo {year} {2017}),\ 10.1038/natrevmats.2017.66}\BibitemShut {NoStop}%
\bibitem [{\citenamefont {Rysaeva}\ \emph {et~al.}(2018)\citenamefont
  {Rysaeva}, \citenamefont {Baimova}, \citenamefont {Lisovenko}, \citenamefont
  {Gorodtsov},\ and\ \citenamefont {Dmitriev}}]{Rysaeva2018}%
  \BibitemOpen
  \bibfield  {author} {\bibinfo {author} {\bibfnamefont {L.~K.}\ \bibnamefont
  {Rysaeva}}, \bibinfo {author} {\bibfnamefont {J.~A.}\ \bibnamefont
  {Baimova}}, \bibinfo {author} {\bibfnamefont {D.~S.}\ \bibnamefont
  {Lisovenko}}, \bibinfo {author} {\bibfnamefont {V.~A.}\ \bibnamefont
  {Gorodtsov}}, \ and\ \bibinfo {author} {\bibfnamefont {S.~V.}\ \bibnamefont
  {Dmitriev}},\ }\href@noop {} {\bibfield  {journal} {\bibinfo  {journal}
  {physica status solidi (b)}\ }\textbf {\bibinfo {volume} {256}},\ \bibinfo
  {pages} {1800049} (\bibinfo {year} {2018})}\BibitemShut {NoStop}%
\bibitem [{\citenamefont {Larsen}, \citenamefont {Signund},\ and\ \citenamefont
  {Bouwsta}(1997)}]{Larsen1997}%
  \BibitemOpen
  \bibfield  {author} {\bibinfo {author} {\bibfnamefont {U.}~\bibnamefont
  {Larsen}}, \bibinfo {author} {\bibfnamefont {O.}~\bibnamefont {Signund}}, \
  and\ \bibinfo {author} {\bibfnamefont {S.}~\bibnamefont {Bouwsta}},\
  }\href@noop {} {\bibfield  {journal} {\bibinfo  {journal} {Journal of
  Microelectromechanical Systems}\ }\textbf {\bibinfo {volume} {6}},\ \bibinfo
  {pages} {99} (\bibinfo {year} {1997})}\BibitemShut {NoStop}%
\bibitem [{\citenamefont {Theocaris}, \citenamefont {Stavroulakis},\ and\
  \citenamefont {Panagiotopoulos}(1997)}]{Theocaris1997}%
  \BibitemOpen
  \bibfield  {author} {\bibinfo {author} {\bibfnamefont {P.~S.}\ \bibnamefont
  {Theocaris}}, \bibinfo {author} {\bibfnamefont {G.~E.}\ \bibnamefont
  {Stavroulakis}}, \ and\ \bibinfo {author} {\bibfnamefont {P.~D.}\
  \bibnamefont {Panagiotopoulos}},\ }\href@noop {} {\bibfield  {journal}
  {\bibinfo  {journal} {Archive of Applied Mechanics (Ingenieur Archiv)}\
  }\textbf {\bibinfo {volume} {67}},\ \bibinfo {pages} {274} (\bibinfo {year}
  {1997})}\BibitemShut {NoStop}%
\bibitem [{\citenamefont {Hanifpour}\ \emph {et~al.}(2018)\citenamefont
  {Hanifpour}, \citenamefont {Petersen}, \citenamefont {Alava},\ and\
  \citenamefont {Zapperi}}]{Hanifpour2018}%
  \BibitemOpen
  \bibfield  {author} {\bibinfo {author} {\bibfnamefont {M.}~\bibnamefont
  {Hanifpour}}, \bibinfo {author} {\bibfnamefont {C.~F.}\ \bibnamefont
  {Petersen}}, \bibinfo {author} {\bibfnamefont {M.~J.}\ \bibnamefont {Alava}},
  \ and\ \bibinfo {author} {\bibfnamefont {S.}~\bibnamefont {Zapperi}},\
  }\href@noop {} {\bibfield  {journal} {\bibinfo  {journal} {The European
  Physical Journal B}\ }\textbf {\bibinfo {volume} {91}} (\bibinfo {year}
  {2018})}\BibitemShut {NoStop}%
\bibitem [{\citenamefont {Reid}\ \emph {et~al.}(2018)\citenamefont {Reid},
  \citenamefont {Pashine}, \citenamefont {Wozniak}, \citenamefont {Jaeger},
  \citenamefont {Liu}, \citenamefont {Nagel},\ and\ \citenamefont
  {de~Pablo}}]{reid2018auxetic}%
  \BibitemOpen
  \bibfield  {author} {\bibinfo {author} {\bibfnamefont {D.~R.}\ \bibnamefont
  {Reid}}, \bibinfo {author} {\bibfnamefont {N.}~\bibnamefont {Pashine}},
  \bibinfo {author} {\bibfnamefont {J.~M.}\ \bibnamefont {Wozniak}}, \bibinfo
  {author} {\bibfnamefont {H.~M.}\ \bibnamefont {Jaeger}}, \bibinfo {author}
  {\bibfnamefont {A.~J.}\ \bibnamefont {Liu}}, \bibinfo {author} {\bibfnamefont
  {S.~R.}\ \bibnamefont {Nagel}}, \ and\ \bibinfo {author} {\bibfnamefont
  {J.~J.}\ \bibnamefont {de~Pablo}},\ }\href@noop {} {\bibfield  {journal}
  {\bibinfo  {journal} {Proceedings of the National Academy of Sciences}\
  }\textbf {\bibinfo {volume} {115}},\ \bibinfo {pages} {E1384} (\bibinfo
  {year} {2018})}\BibitemShut {NoStop}%
\bibitem [{\citenamefont {Dong}, \citenamefont {Stone},\ and\ \citenamefont
  {Lakes}(2010)}]{Dong2010}%
  \BibitemOpen
  \bibfield  {author} {\bibinfo {author} {\bibfnamefont {L.}~\bibnamefont
  {Dong}}, \bibinfo {author} {\bibfnamefont {D.~S.}\ \bibnamefont {Stone}}, \
  and\ \bibinfo {author} {\bibfnamefont {R.~S.}\ \bibnamefont {Lakes}},\
  }\href@noop {} {\bibfield  {journal} {\bibinfo  {journal} {Philosophical
  Magazine Letters}\ }\textbf {\bibinfo {volume} {90}},\ \bibinfo {pages} {23}
  (\bibinfo {year} {2010})}\BibitemShut {NoStop}%
\bibitem [{\citenamefont {Kou}\ \emph {et~al.}(2016)\citenamefont {Kou},
  \citenamefont {Ma}, \citenamefont {Tang}, \citenamefont {Sun}, \citenamefont
  {Du},\ and\ \citenamefont {Chen}}]{Kou2016}%
  \BibitemOpen
  \bibfield  {author} {\bibinfo {author} {\bibfnamefont {L.}~\bibnamefont
  {Kou}}, \bibinfo {author} {\bibfnamefont {Y.}~\bibnamefont {Ma}}, \bibinfo
  {author} {\bibfnamefont {C.}~\bibnamefont {Tang}}, \bibinfo {author}
  {\bibfnamefont {Z.}~\bibnamefont {Sun}}, \bibinfo {author} {\bibfnamefont
  {A.}~\bibnamefont {Du}}, \ and\ \bibinfo {author} {\bibfnamefont
  {C.}~\bibnamefont {Chen}},\ }\href@noop {} {\bibfield  {journal} {\bibinfo
  {journal} {Nano Letters}\ }\textbf {\bibinfo {volume} {16}},\ \bibinfo
  {pages} {7910} (\bibinfo {year} {2016})}\BibitemShut {NoStop}%
\bibitem [{\citenamefont {Hirotsu}(1991)}]{Hirotsu1991}%
  \BibitemOpen
  \bibfield  {author} {\bibinfo {author} {\bibfnamefont {S.}~\bibnamefont
  {Hirotsu}},\ }\href {\doibase 10.1063/1.460672} {\bibfield  {journal}
  {\bibinfo  {journal} {The Journal of Chemical Physics}\ }\textbf {\bibinfo
  {volume} {94}},\ \bibinfo {pages} {3949} (\bibinfo {year}
  {1991})}\BibitemShut {NoStop}%
\bibitem [{\citenamefont {Boon}\ and\ \citenamefont
  {Schurtenberger}(2017)}]{Boon2017}%
  \BibitemOpen
  \bibfield  {author} {\bibinfo {author} {\bibfnamefont {N.}~\bibnamefont
  {Boon}}\ and\ \bibinfo {author} {\bibfnamefont {P.}~\bibnamefont
  {Schurtenberger}},\ }\href {\doibase 10.1039/c7cp02434g} {\bibfield
  {journal} {\bibinfo  {journal} {Physical Chemistry Chemical Physics}\
  }\textbf {\bibinfo {volume} {19}},\ \bibinfo {pages} {23740} (\bibinfo {year}
  {2017})}\BibitemShut {NoStop}%
\bibitem [{\citenamefont {Ninarello}, \citenamefont {Ruiz-Franco},\ and\
  \citenamefont {Zaccarelli}(2022)}]{Ninarello2022}%
  \BibitemOpen
  \bibfield  {author} {\bibinfo {author} {\bibfnamefont {A.}~\bibnamefont
  {Ninarello}}, \bibinfo {author} {\bibfnamefont {J.}~\bibnamefont
  {Ruiz-Franco}}, \ and\ \bibinfo {author} {\bibfnamefont {E.}~\bibnamefont
  {Zaccarelli}},\ }\href {\doibase 10.1038/s41467-022-28026-z} {\bibfield
  {journal} {\bibinfo  {journal} {Nature Communications}\ }\textbf {\bibinfo
  {volume} {13}} (\bibinfo {year} {2022}),\
  10.1038/s41467-022-28026-z}\BibitemShut {NoStop}%
\bibitem [{\citenamefont {Doghri}(2013)}]{Doghri2013}%
  \BibitemOpen
  \bibfield  {author} {\bibinfo {author} {\bibfnamefont {I.}~\bibnamefont
  {Doghri}},\ }\href@noop {} {\emph {\bibinfo {title} {Mechanics of deformable
  solids: Linear, Nonlinear, Analytical and Computational Aspects}}}\ (\bibinfo
   {publisher} {Springer},\ \bibinfo {year} {2013})\BibitemShut {NoStop}%
\bibitem [{\citenamefont {Rovigatti}\ \emph {et~al.}(2019)\citenamefont
  {Rovigatti}, \citenamefont {Gnan}, \citenamefont {Ninarello},\ and\
  \citenamefont {Zaccarelli}}]{Rovigatti2019}%
  \BibitemOpen
  \bibfield  {author} {\bibinfo {author} {\bibfnamefont {L.}~\bibnamefont
  {Rovigatti}}, \bibinfo {author} {\bibfnamefont {N.}~\bibnamefont {Gnan}},
  \bibinfo {author} {\bibfnamefont {A.}~\bibnamefont {Ninarello}}, \ and\
  \bibinfo {author} {\bibfnamefont {E.}~\bibnamefont {Zaccarelli}},\ }\href
  {\doibase 10.1021/acs.macromol.9b00099} {\bibfield  {journal} {\bibinfo
  {journal} {Macromolecules}\ }\textbf {\bibinfo {volume} {52}},\ \bibinfo
  {pages} {4895} (\bibinfo {year} {2019})}\BibitemShut {NoStop}%
\bibitem [{\citenamefont {Grest}\ and\ \citenamefont
  {Kremer}(1986)}]{Grest1986}%
  \BibitemOpen
  \bibfield  {author} {\bibinfo {author} {\bibfnamefont {G.~S.}\ \bibnamefont
  {Grest}}\ and\ \bibinfo {author} {\bibfnamefont {K.}~\bibnamefont {Kremer}},\
  }\href@noop {} {\bibfield  {journal} {\bibinfo  {journal} {Physical Review
  A}\ }\textbf {\bibinfo {volume} {33}},\ \bibinfo {pages} {3628} (\bibinfo
  {year} {1986})}\BibitemShut {NoStop}%
\bibitem [{\citenamefont {Kremer}\ and\ \citenamefont
  {Grest}(1990)}]{Kremer1990}%
  \BibitemOpen
  \bibfield  {author} {\bibinfo {author} {\bibfnamefont {K.}~\bibnamefont
  {Kremer}}\ and\ \bibinfo {author} {\bibfnamefont {G.~S.}\ \bibnamefont
  {Grest}},\ }\href@noop {} {\bibfield  {journal} {\bibinfo  {journal} {The
  Journal of Chemical Physics}\ }\textbf {\bibinfo {volume} {92}},\ \bibinfo
  {pages} {5057} (\bibinfo {year} {1990})}\BibitemShut {NoStop}%
\bibitem [{\citenamefont {Duering}, \citenamefont {Kremer},\ and\ \citenamefont
  {Grest}(1992)}]{Duering1992}%
  \BibitemOpen
  \bibfield  {author} {\bibinfo {author} {\bibfnamefont {E.}~\bibnamefont
  {Duering}}, \bibinfo {author} {\bibfnamefont {K.}~\bibnamefont {Kremer}}, \
  and\ \bibinfo {author} {\bibfnamefont {G.}~\bibnamefont {Grest}},\ }in\
  \href@noop {} {\emph {\bibinfo {booktitle} {Physics of Polymer Networks}}}\
  (\bibinfo  {publisher} {Springer},\ \bibinfo {year} {1992})\ pp.\ \bibinfo
  {pages} {13--15}\BibitemShut {NoStop}%
\bibitem [{\citenamefont {Duering}, \citenamefont {Kremer},\ and\ \citenamefont
  {Grest}(1994)}]{Duering1994}%
  \BibitemOpen
  \bibfield  {author} {\bibinfo {author} {\bibfnamefont {E.~R.}\ \bibnamefont
  {Duering}}, \bibinfo {author} {\bibfnamefont {K.}~\bibnamefont {Kremer}}, \
  and\ \bibinfo {author} {\bibfnamefont {G.~S.}\ \bibnamefont {Grest}},\
  }\href@noop {} {\bibfield  {journal} {\bibinfo  {journal} {The Journal of
  Chemical Physics}\ }\textbf {\bibinfo {volume} {101}},\ \bibinfo {pages}
  {8169} (\bibinfo {year} {1994})}\BibitemShut {NoStop}%
\bibitem [{\citenamefont {Kenkare}\ \emph {et~al.}(1998)\citenamefont
  {Kenkare}, \citenamefont {Smith}, \citenamefont {Hall},\ and\ \citenamefont
  {Khan}}]{Kenkare1998}%
  \BibitemOpen
  \bibfield  {author} {\bibinfo {author} {\bibfnamefont {N.}~\bibnamefont
  {Kenkare}}, \bibinfo {author} {\bibfnamefont {S.}~\bibnamefont {Smith}},
  \bibinfo {author} {\bibfnamefont {C.}~\bibnamefont {Hall}}, \ and\ \bibinfo
  {author} {\bibfnamefont {S.}~\bibnamefont {Khan}},\ }\href@noop {} {\bibfield
   {journal} {\bibinfo  {journal} {Macromolecules}\ }\textbf {\bibinfo {volume}
  {31}},\ \bibinfo {pages} {5861} (\bibinfo {year} {1998})}\BibitemShut
  {NoStop}%
\bibitem [{\citenamefont {Auhl}\ \emph {et~al.}(2003)\citenamefont {Auhl},
  \citenamefont {Everaers}, \citenamefont {Grest}, \citenamefont {Kremer},\
  and\ \citenamefont {Plimpton}}]{Auhl2003}%
  \BibitemOpen
  \bibfield  {author} {\bibinfo {author} {\bibfnamefont {R.}~\bibnamefont
  {Auhl}}, \bibinfo {author} {\bibfnamefont {R.}~\bibnamefont {Everaers}},
  \bibinfo {author} {\bibfnamefont {G.~S.}\ \bibnamefont {Grest}}, \bibinfo
  {author} {\bibfnamefont {K.}~\bibnamefont {Kremer}}, \ and\ \bibinfo {author}
  {\bibfnamefont {S.~J.}\ \bibnamefont {Plimpton}},\ }\href@noop {} {\bibfield
  {journal} {\bibinfo  {journal} {The Journal of Chemical Physics}\ }\textbf
  {\bibinfo {volume} {119}},\ \bibinfo {pages} {12718} (\bibinfo {year}
  {2003})}\BibitemShut {NoStop}%
\bibitem [{\citenamefont {Everaers}\ \emph {et~al.}(2004)\citenamefont
  {Everaers}, \citenamefont {Sukumaran}, \citenamefont {Grest}, \citenamefont
  {Svaneborg}, \citenamefont {Sivasubramanian},\ and\ \citenamefont
  {Kremer}}]{Everaers2004}%
  \BibitemOpen
  \bibfield  {author} {\bibinfo {author} {\bibfnamefont {R.}~\bibnamefont
  {Everaers}}, \bibinfo {author} {\bibfnamefont {S.~K.}\ \bibnamefont
  {Sukumaran}}, \bibinfo {author} {\bibfnamefont {G.~S.}\ \bibnamefont
  {Grest}}, \bibinfo {author} {\bibfnamefont {C.}~\bibnamefont {Svaneborg}},
  \bibinfo {author} {\bibfnamefont {A.}~\bibnamefont {Sivasubramanian}}, \ and\
  \bibinfo {author} {\bibfnamefont {K.}~\bibnamefont {Kremer}},\ }\href@noop {}
  {\bibfield  {journal} {\bibinfo  {journal} {Science}\ }\textbf {\bibinfo
  {volume} {303}},\ \bibinfo {pages} {823} (\bibinfo {year}
  {2004})}\BibitemShut {NoStop}%
\bibitem [{\citenamefont {Lang}(2013)}]{Lang2013}%
  \BibitemOpen
  \bibfield  {author} {\bibinfo {author} {\bibfnamefont {M.}~\bibnamefont
  {Lang}},\ }\href@noop {} {\bibfield  {journal} {\bibinfo  {journal}
  {Macromolecules}\ }\textbf {\bibinfo {volume} {46}},\ \bibinfo {pages} {9782}
  (\bibinfo {year} {2013})}\BibitemShut {NoStop}%
\bibitem [{\citenamefont {Duering}, \citenamefont {Kremer},\ and\ \citenamefont
  {Grest}(1991)}]{Duering1991}%
  \BibitemOpen
  \bibfield  {author} {\bibinfo {author} {\bibfnamefont {E.~R.}\ \bibnamefont
  {Duering}}, \bibinfo {author} {\bibfnamefont {K.}~\bibnamefont {Kremer}}, \
  and\ \bibinfo {author} {\bibfnamefont {G.~S.}\ \bibnamefont {Grest}},\ }\href
  {\doibase 10.1103/physrevlett.67.3531} {\bibfield  {journal} {\bibinfo
  {journal} {Physical Review Letters}\ }\textbf {\bibinfo {volume} {67}},\
  \bibinfo {pages} {3531} (\bibinfo {year} {1991})}\BibitemShut {NoStop}%
\bibitem [{\citenamefont {Gnan}\ \emph {et~al.}(2017)\citenamefont {Gnan},
  \citenamefont {Rovigatti}, \citenamefont {Bergman},\ and\ \citenamefont
  {Zaccarelli}}]{Gnan2017}%
  \BibitemOpen
  \bibfield  {author} {\bibinfo {author} {\bibfnamefont {N.}~\bibnamefont
  {Gnan}}, \bibinfo {author} {\bibfnamefont {L.}~\bibnamefont {Rovigatti}},
  \bibinfo {author} {\bibfnamefont {M.}~\bibnamefont {Bergman}}, \ and\
  \bibinfo {author} {\bibfnamefont {E.}~\bibnamefont {Zaccarelli}},\ }\href
  {\doibase 10.1021/acs.macromol.7b01600} {\bibfield  {journal} {\bibinfo
  {journal} {Macromolecules}\ }\textbf {\bibinfo {volume} {50}},\ \bibinfo
  {pages} {8777} (\bibinfo {year} {2017})}\BibitemShut {NoStop}%
\bibitem [{\citenamefont {Sorichetti}\ \emph {et~al.}(2021)\citenamefont
  {Sorichetti}, \citenamefont {Ninarello}, \citenamefont {Ruiz-Franco},
  \citenamefont {Hugouvieux}, \citenamefont {Kob}, \citenamefont {Zaccarelli},\
  and\ \citenamefont {Rovigatti}}]{Sorichetti2021}%
  \BibitemOpen
  \bibfield  {author} {\bibinfo {author} {\bibfnamefont {V.}~\bibnamefont
  {Sorichetti}}, \bibinfo {author} {\bibfnamefont {A.}~\bibnamefont
  {Ninarello}}, \bibinfo {author} {\bibfnamefont {J.~M.}\ \bibnamefont
  {Ruiz-Franco}}, \bibinfo {author} {\bibfnamefont {V.}~\bibnamefont
  {Hugouvieux}}, \bibinfo {author} {\bibfnamefont {W.}~\bibnamefont {Kob}},
  \bibinfo {author} {\bibfnamefont {E.}~\bibnamefont {Zaccarelli}}, \ and\
  \bibinfo {author} {\bibfnamefont {L.}~\bibnamefont {Rovigatti}},\ }\href@noop
  {} {\bibfield  {journal} {\bibinfo  {journal} {Macromolecules}\ }\textbf
  {\bibinfo {volume} {54}},\ \bibinfo {pages} {3769} (\bibinfo {year}
  {2021})}\BibitemShut {NoStop}%
\bibitem [{\citenamefont {Sorichetti}\ \emph {et~al.}(2023)\citenamefont
  {Sorichetti}, \citenamefont {Ninarello}, \citenamefont {Ruiz-Franco},
  \citenamefont {Hugouvieux}, \citenamefont {Zaccarelli}, \citenamefont
  {Micheletti}, \citenamefont {Kob},\ and\ \citenamefont
  {Rovigatti}}]{Sorichetti2023}%
  \BibitemOpen
  \bibfield  {author} {\bibinfo {author} {\bibfnamefont {V.}~\bibnamefont
  {Sorichetti}}, \bibinfo {author} {\bibfnamefont {A.}~\bibnamefont
  {Ninarello}}, \bibinfo {author} {\bibfnamefont {J.}~\bibnamefont
  {Ruiz-Franco}}, \bibinfo {author} {\bibfnamefont {V.}~\bibnamefont
  {Hugouvieux}}, \bibinfo {author} {\bibfnamefont {E.}~\bibnamefont
  {Zaccarelli}}, \bibinfo {author} {\bibfnamefont {C.}~\bibnamefont
  {Micheletti}}, \bibinfo {author} {\bibfnamefont {W.}~\bibnamefont {Kob}}, \
  and\ \bibinfo {author} {\bibfnamefont {L.}~\bibnamefont {Rovigatti}},\ }\href
  {\doibase 10.1063/5.0134271} {\bibfield  {journal} {\bibinfo  {journal} {The
  Journal of Chemical Physics}\ }\textbf {\bibinfo {volume} {158}},\ \bibinfo
  {pages} {074905} (\bibinfo {year} {2023})}\BibitemShut {NoStop}%
\bibitem [{\citenamefont {Sciortino}\ \emph {et~al.}(2007)\citenamefont
  {Sciortino}, \citenamefont {Bianchi}, \citenamefont {Douglas},\ and\
  \citenamefont {Tartaglia}}]{Sciortino2007}%
  \BibitemOpen
  \bibfield  {author} {\bibinfo {author} {\bibfnamefont {F.}~\bibnamefont
  {Sciortino}}, \bibinfo {author} {\bibfnamefont {E.}~\bibnamefont {Bianchi}},
  \bibinfo {author} {\bibfnamefont {J.~F.}\ \bibnamefont {Douglas}}, \ and\
  \bibinfo {author} {\bibfnamefont {P.}~\bibnamefont {Tartaglia}},\ }\href
  {\doibase 10.1063/1.2730797} {\bibfield  {journal} {\bibinfo  {journal} {The
  Journal of Chemical Physics}\ }\textbf {\bibinfo {volume} {126}},\ \bibinfo
  {pages} {194903} (\bibinfo {year} {2007})}\BibitemShut {NoStop}%
\bibitem [{\citenamefont {Sciortino}\ and\ \citenamefont
  {Zaccarelli}(2011)}]{Sciortino2011}%
  \BibitemOpen
  \bibfield  {author} {\bibinfo {author} {\bibfnamefont {F.}~\bibnamefont
  {Sciortino}}\ and\ \bibinfo {author} {\bibfnamefont {E.}~\bibnamefont
  {Zaccarelli}},\ }\href@noop {} {\bibfield  {journal} {\bibinfo  {journal}
  {Current Opinion in Solid State and Materials Science}\ }\textbf {\bibinfo
  {volume} {15}},\ \bibinfo {pages} {246} (\bibinfo {year} {2011})}\BibitemShut
  {NoStop}%
\bibitem [{\citenamefont {Rovigatti}\ \emph {et~al.}(2014)\citenamefont
  {Rovigatti}, \citenamefont {{\v{S}}ulc}, \citenamefont {Reguly},\ and\
  \citenamefont {Romano}}]{Rovigatti2014}%
  \BibitemOpen
  \bibfield  {author} {\bibinfo {author} {\bibfnamefont {L.}~\bibnamefont
  {Rovigatti}}, \bibinfo {author} {\bibfnamefont {P.}~\bibnamefont
  {{\v{S}}ulc}}, \bibinfo {author} {\bibfnamefont {I.~Z.}\ \bibnamefont
  {Reguly}}, \ and\ \bibinfo {author} {\bibfnamefont {F.}~\bibnamefont
  {Romano}},\ }\href {\doibase 10.1002/jcc.23763} {\bibfield  {journal}
  {\bibinfo  {journal} {Journal of Computational Chemistry}\ }\textbf {\bibinfo
  {volume} {36}},\ \bibinfo {pages} {1} (\bibinfo {year} {2014})}\BibitemShut
  {NoStop}%
\bibitem [{\citenamefont {Poppleton}\ \emph {et~al.}(2023)\citenamefont
  {Poppleton}, \citenamefont {Matthies}, \citenamefont {Mandal}, \citenamefont
  {Romano}, \citenamefont {Šulc},\ and\ \citenamefont
  {Rovigatti}}]{Poppleton2023}%
  \BibitemOpen
  \bibfield  {author} {\bibinfo {author} {\bibfnamefont {E.}~\bibnamefont
  {Poppleton}}, \bibinfo {author} {\bibfnamefont {M.}~\bibnamefont {Matthies}},
  \bibinfo {author} {\bibfnamefont {D.}~\bibnamefont {Mandal}}, \bibinfo
  {author} {\bibfnamefont {F.}~\bibnamefont {Romano}}, \bibinfo {author}
  {\bibfnamefont {P.}~\bibnamefont {Šulc}}, \ and\ \bibinfo {author}
  {\bibfnamefont {L.}~\bibnamefont {Rovigatti}},\ }\href {\doibase
  10.21105/joss.04693} {\bibfield  {journal} {\bibinfo  {journal} {Journal of
  Open Source Software}\ }\textbf {\bibinfo {volume} {8}},\ \bibinfo {pages}
  {4693} (\bibinfo {year} {2023})}\BibitemShut {NoStop}%
\bibitem [{\citenamefont {Ninarello}\ \emph {et~al.}(2019)\citenamefont
  {Ninarello}, \citenamefont {Crassous}, \citenamefont {Paloli}, \citenamefont
  {Camerin}, \citenamefont {Gnan}, \citenamefont {Rovigatti}, \citenamefont
  {Schurtenberger},\ and\ \citenamefont {Zaccarelli}}]{Ninarello2019}%
  \BibitemOpen
  \bibfield  {author} {\bibinfo {author} {\bibfnamefont {A.}~\bibnamefont
  {Ninarello}}, \bibinfo {author} {\bibfnamefont {J.~J.}\ \bibnamefont
  {Crassous}}, \bibinfo {author} {\bibfnamefont {D.}~\bibnamefont {Paloli}},
  \bibinfo {author} {\bibfnamefont {F.}~\bibnamefont {Camerin}}, \bibinfo
  {author} {\bibfnamefont {N.}~\bibnamefont {Gnan}}, \bibinfo {author}
  {\bibfnamefont {L.}~\bibnamefont {Rovigatti}}, \bibinfo {author}
  {\bibfnamefont {P.}~\bibnamefont {Schurtenberger}}, \ and\ \bibinfo {author}
  {\bibfnamefont {E.}~\bibnamefont {Zaccarelli}},\ }\href {\doibase
  10.1021/acs.macromol.9b01122} {\bibfield  {journal} {\bibinfo  {journal}
  {Macromolecules}\ }\textbf {\bibinfo {volume} {52}},\ \bibinfo {pages} {7584}
  (\bibinfo {year} {2019})}\BibitemShut {NoStop}%
\bibitem [{\citenamefont {Plimpton}(1995)}]{Plimpton1995}%
  \BibitemOpen
  \bibfield  {author} {\bibinfo {author} {\bibfnamefont {S.}~\bibnamefont
  {Plimpton}},\ }\href@noop {} {\bibfield  {journal} {\bibinfo  {journal}
  {Journal of Computational Physics}\ }\textbf {\bibinfo {volume} {117}},\
  \bibinfo {pages} {1} (\bibinfo {year} {1995})}\BibitemShut {NoStop}%
\bibitem [{\citenamefont {Rivas-Barbosa}\ \emph {et~al.}(2022)\citenamefont
  {Rivas-Barbosa}, \citenamefont {Ruiz-Franco}, \citenamefont
  {Lara-Pe{\~{n}}a}, \citenamefont {Cardellini}, \citenamefont
  {Licea-Claverie}, \citenamefont {Camerin}, \citenamefont {Zaccarelli},\ and\
  \citenamefont {Laurati}}]{RivasBarbosa2022}%
  \BibitemOpen
  \bibfield  {author} {\bibinfo {author} {\bibfnamefont {R.}~\bibnamefont
  {Rivas-Barbosa}}, \bibinfo {author} {\bibfnamefont {J.}~\bibnamefont
  {Ruiz-Franco}}, \bibinfo {author} {\bibfnamefont {M.~A.}\ \bibnamefont
  {Lara-Pe{\~{n}}a}}, \bibinfo {author} {\bibfnamefont {J.}~\bibnamefont
  {Cardellini}}, \bibinfo {author} {\bibfnamefont {A.}~\bibnamefont
  {Licea-Claverie}}, \bibinfo {author} {\bibfnamefont {F.}~\bibnamefont
  {Camerin}}, \bibinfo {author} {\bibfnamefont {E.}~\bibnamefont {Zaccarelli}},
  \ and\ \bibinfo {author} {\bibfnamefont {M.}~\bibnamefont {Laurati}},\ }\href
  {\doibase 10.1021/acs.macromol.1c02171} {\bibfield  {journal} {\bibinfo
  {journal} {Macromolecules}\ }\textbf {\bibinfo {volume} {55}},\ \bibinfo
  {pages} {1834} (\bibinfo {year} {2022})}\BibitemShut {NoStop}%
\bibitem [{\citenamefont {Camerin}\ \emph {et~al.}(2020)\citenamefont
  {Camerin}, \citenamefont {Gnan}, \citenamefont {Ruiz-Franco}, \citenamefont
  {Ninarello}, \citenamefont {Rovigatti},\ and\ \citenamefont
  {Zaccarelli}}]{Camerin2020}%
  \BibitemOpen
  \bibfield  {author} {\bibinfo {author} {\bibfnamefont {F.}~\bibnamefont
  {Camerin}}, \bibinfo {author} {\bibfnamefont {N.}~\bibnamefont {Gnan}},
  \bibinfo {author} {\bibfnamefont {J.}~\bibnamefont {Ruiz-Franco}}, \bibinfo
  {author} {\bibfnamefont {A.}~\bibnamefont {Ninarello}}, \bibinfo {author}
  {\bibfnamefont {L.}~\bibnamefont {Rovigatti}}, \ and\ \bibinfo {author}
  {\bibfnamefont {E.}~\bibnamefont {Zaccarelli}},\ }\href {\doibase
  10.1103/physrevx.10.031012} {\bibfield  {journal} {\bibinfo  {journal}
  {Physical Review X}\ }\textbf {\bibinfo {volume} {10}} (\bibinfo {year}
  {2020}),\ 10.1103/physrevx.10.031012}\BibitemShut {NoStop}%
\bibitem [{\citenamefont {Little}\ \emph {et~al.}(2023)\citenamefont {Little},
  \citenamefont {Levine}, \citenamefont {Singh},\ and\ \citenamefont
  {Bruinsma}}]{Little2023}%
  \BibitemOpen
  \bibfield  {author} {\bibinfo {author} {\bibfnamefont {J.}~\bibnamefont
  {Little}}, \bibinfo {author} {\bibfnamefont {A.~J.}\ \bibnamefont {Levine}},
  \bibinfo {author} {\bibfnamefont {A.~R.}\ \bibnamefont {Singh}}, \ and\
  \bibinfo {author} {\bibfnamefont {R.}~\bibnamefont {Bruinsma}},\ }\href
  {\doibase 10.1103/physreve.107.024418} {\bibfield  {journal} {\bibinfo
  {journal} {Physical Review E}\ }\textbf {\bibinfo {volume} {107}} (\bibinfo
  {year} {2023}),\ 10.1103/physreve.107.024418}\BibitemShut {NoStop}%
\bibitem [{\citenamefont {Aggarwal}\ \emph {et~al.}(2016)\citenamefont
  {Aggarwal}, \citenamefont {May}, \citenamefont {Brooks},\ and\ \citenamefont
  {Klug}}]{Aggarwal2016}%
  \BibitemOpen
  \bibfield  {author} {\bibinfo {author} {\bibfnamefont {A.}~\bibnamefont
  {Aggarwal}}, \bibinfo {author} {\bibfnamefont {E.~R.}\ \bibnamefont {May}},
  \bibinfo {author} {\bibfnamefont {C.~L.}\ \bibnamefont {Brooks}}, \ and\
  \bibinfo {author} {\bibfnamefont {W.~S.}\ \bibnamefont {Klug}},\ }\href
  {\doibase 10.1103/physreve.93.012417} {\bibfield  {journal} {\bibinfo
  {journal} {Physical Review E}\ }\textbf {\bibinfo {volume} {93}} (\bibinfo
  {year} {2016}),\ 10.1103/physreve.93.012417}\BibitemShut {NoStop}%
\bibitem [{\citenamefont {Aggarwal}(2018)}]{Aggarwal2018}%
  \BibitemOpen
  \bibfield  {author} {\bibinfo {author} {\bibfnamefont {A.}~\bibnamefont
  {Aggarwal}},\ }\href {\doibase 10.1103/physreve.97.032414} {\bibfield
  {journal} {\bibinfo  {journal} {Physical Review E}\ }\textbf {\bibinfo
  {volume} {97}} (\bibinfo {year} {2018}),\
  10.1103/physreve.97.032414}\BibitemShut {NoStop}%
\bibitem [{\citenamefont {Landau}\ \emph {et~al.}(2009)\citenamefont {Landau},
  \citenamefont {Lifšic}, \citenamefont {Landau},\ and\ \citenamefont
  {Landau}}]{landau_theory_2009}%
  \BibitemOpen
  \bibfield  {author} {\bibinfo {author} {\bibfnamefont {L.~D.}\ \bibnamefont
  {Landau}}, \bibinfo {author} {\bibfnamefont {E.~M.}\ \bibnamefont {Lifšic}},
  \bibinfo {author} {\bibfnamefont {L.~D.}\ \bibnamefont {Landau}}, \ and\
  \bibinfo {author} {\bibfnamefont {L.~D.}\ \bibnamefont {Landau}},\
  }\href@noop {} {\emph {\bibinfo {title} {Theory of elasticity}}},\ \bibinfo
  {edition} {3rd}\ ed.,\ \bibinfo {series} {Course of theoretical physics /
  {L}. {D}. {Landau} and {E}. {M}. {Lifshitz}}\ No.~\bibinfo {number} {7}\
  (\bibinfo  {publisher} {Elsevier, Butterworth-Heinemann},\ \bibinfo {address}
  {Amsterdam Heidelberg},\ \bibinfo {year} {2009})\BibitemShut {NoStop}%
\bibitem [{\citenamefont {Rubinstein}(2014)}]{Rubinstein2014}%
  \BibitemOpen
  \bibfield  {author} {\bibinfo {author} {\bibfnamefont {M.}~\bibnamefont
  {Rubinstein}},\ }\href@noop {} {\emph {\bibinfo {title} {Polymer physics}}}\
  (\bibinfo  {publisher} {Oxford University Press},\ \bibinfo {year}
  {2014})\BibitemShut {NoStop}%
\end{thebibliography}%

\end{document}